\documentclass[12pt]{article}
\usepackage{latexsym}
\usepackage{epsfig,amssymb,euscript}
\usepackage{amsmath}
\numberwithin{equation}{section}  

\newsavebox{\ns}
\newsavebox{\dbrane}
\newsavebox{\dbshort}

\def\be{\begin{equation}}
\def\ee{\end{equation}}
\def\bea{\begin{eqnarray}}
\def\eea{\end{eqnarray}}

\newcommand{\nn}{\nonumber}

\newcommand{\eqn}[1]{(\ref{#1})}

\def\Dslash{\,\,{\raise.15ex\hbox{/}\mkern-12mu D}}
\def\Dbarslash{\,\,{\raise.15ex\hbox{/}\mkern-12mu {\bar D}}}
\def\delslash{\,\,{\raise.15ex\hbox{/}\mkern-9mu \partial}}
\def\delbarslash{\,\,{\raise.15ex\hbox{/}\mkern-9mu {\bar\partial}}}
\def\pslash{\,\,{\raise.15ex\hbox{/}\mkern-9mu p}}
\def\calDslash{\,\,{\raise.15ex\hbox{/}\mkern-12mu {\cal D}}}

\newcommand\R{\mathbb{R}}
\newcommand\Z{\mathbb{Z}}

\newcommand\C{\mathbb{C}}

\newcommand\diff{\mathrm{d}}
\newcommand{\de}{\partial}

\newcommand{\vol}{\mathrm{vol}}
\newcommand{\eab}{\epsilon_{\alpha\beta}}

\newcommand{\qed}{\nobreak \ifvmode \relax \else \ifdim\lastskip<1.5em \hskip-\lastskip \hskip1.5em plus0em minus0.5em \fi \nobreak \vrule height0.75em width0.5em depth0.25em\fi}

\begin{document}
\begin{titlepage}
\begin{center}
\today

\vskip 2 cm {\Large \bf Baryonic branches and resolutions}\\[5mm]
{\Large\bf of Ricci-flat K\"ahler cones}\\[6mm]

\vskip 1cm
{Dario Martelli$^{1*}$ and James Sparks$^{2}$}\\
\vskip 1 cm

1: {\em Institute for Advanced Study\\
Einstein Drive, Princeton, NJ 08540, U.S.A.}\\
\vskip 0.5cm
2: {\em Mathematical Institute, University of Oxford,\\
 24-29 St Giles', Oxford OX1 3LB, U.K.}\\

\vskip 2cm

\end{center}

\begin{abstract}
\noindent

We consider deformations of $\mathcal{N}=1$
superconformal field theories that are AdS/CFT 
dual to Type IIB
string theory on Sasaki-Einstein manifolds, characterised by
non-zero vacuum expectation values for certain baryonic operators. 
Such baryonic branches are constructed from (partially)
resolved, asymptotically conical Ricci-flat K\"ahler manifolds,
together with a choice of point where the stack of D3-branes is
placed. The complete solution then describes a renormalisation group
flow between two AdS fixed points. We discuss the use of probe
Euclidean D3-branes in these backgrounds as a means to
compute expectation values of baryonic operators. The
$Y^{p,q}$ theories are used as illustrative examples throughout the
paper. In particular, we present supergravity solutions
describing flows from the $Y^{p,q}$ theories to various different 
orbifold field theories in the infra-red, and successfully match this to an
explicit field theory analysis. 
\end{abstract}

\vfill
\hrule width 5cm
\vskip 5mm

{\noindent $^*$ {\small On leave from: \emph{Blackett Laboratory,
Imperial College, London SW7 2AZ, U.K.}}}

\end{titlepage}
\pagestyle{plain}
\setcounter{page}{1}
\newcounter{bean}
\baselineskip18pt

\tableofcontents


\section{Introduction}

The AdS/CFT correspondence \cite{Maldacena:1997re} may be used as a powerful tool for
addressing difficult problems in field
theory using geometric techniques. The correspondence provides us with a precise map between
a large class of conformal field theories, together with certain deformations of these theories, and various
types of geometry. A rich set of examples consists of Type IIB string theory in the
background AdS$_5\times Y$, where $Y$ is a Sasaki-Einstein five-manifold \cite{Kehagias:1998gn,KW,Acharya:1998db,MP}.
 For example, one may take $Y=T^{1,1}$ \cite{KW}, or the more recently discovered infinite families
of Sasaki-Einstein manifolds, $Y^{p,q}$
\cite{paper1,paper2} and $L^{a,b,c}$ \cite{Cvetic:2005vk,Labc}.
In all these cases, the dual field theories
\cite{toricpaper,Bertolini:2004xf,quiverpaper,Benvenuti:2005ja,tilings,Butti:2005sw} are conjectured to be
supersymmetric gauge theories, at an infra-red (IR) conformal fixed
point of the renormalisation group (RG). More briefly, they are ${\cal N}=1$ SCFTs.

Such AdS$_5$ backgrounds arise from placing a large number $N$
 of parallel D3-branes at the singular point of a Calabi-Yau singularity $C(Y)$, equipped with a Ricci-flat
 K\"ahler cone metric
 \bea
 g_{C(Y)} & = & \diff r^2 + r^2 g_Y~.
 \label{firstmetric}
 \eea
The backreaction of the branes induces a warp factor, which is essentially the Green's
function for the metric (\ref{firstmetric}), and produces an AdS$_5\times Y$ geometry
together with $N$ units of Ramond-Ramond (RR) five-form flux.

One  interesting generalisation of the original AdS/CFT
correspondence is to consider deformations of the conformal field
theories and their dual geometric description. The class of
deformations that we will study in this paper correspond to giving
vacuum expectation values (VEVs) to certain \emph{baryonic
operators}. These types of deformation allow one to explore
different baryonic branches of the moduli space of a given theory,
and are in general related to (partial) \emph{resolutions} of the
conical Calabi-Yau singularity. In the context of the
conifold theory \cite{KW} some features of
these solutions were discussed in \cite{KW2}, and recently expanded
upon\footnote{For other examples, see
\cite{Benvenuti:2005qb,Chen:2007em,Cvetic:2007nv}.} in \cite{KM}.
However, a systematic discussion of these baryonic branches, from an
AdS/CFT perspective, has not appeared before. The full
ten-dimensional metric is simply a warped product \bea g_{10} & = &
H^{-1/2} g_{\mathbb{R}^{1,3}}+ H^{1/2} g_X~, \eea where $g_X$ is a
Ricci-flat K\"ahler metric that is asymptotic to the conical metric
\eqn{firstmetric}, and the warp factor $H$ is the Green's function
on $X$, sourced by a  stack of D3-branes that are localised at some
point $\mathbf{x}_0\in X$. 
The baryonic branches
considered here are different from the kind studied in
\cite{Butti:2004pk,Dymarsky:2005xt}, where the field theory
undergoes a cascade of Seiberg dualities. Nevertheless, the results presented 
in this paper may be useful  for obtaining a better understanding of baryonic deformations 
of non-conformal theories  as well.

Until recently,
explicit Ricci-flat K\"ahler metrics of this kind were not known, apart from the case of the conifold
and its $\Z_2$ orbifold \cite{Pando Zayas:2001iw}\footnote{More generally one may also study
the Ricci-flat K\"ahler metrics on the canonical line bundles over
K\"ahler-Einstein manifolds constructed in \cite{BB,PP}, which are explicit up to the K\"ahler-Einstein metric.}.
In \cite{np1} we presented families of explicit
Ricci-flat K\"ahler partial resolutions of conical singularities in all dimensions. These included
several classes in three complex dimensions that give rise to toric partial resolutions of the $Y^{p,q}$
singularities (see also \cite{japanese,lupope,Balasubramanian:2007dv}). In the present paper we will
further discuss these metrics, providing their toric geometry description and their dual gauge theory
interpretation. In fact, these are just examples of a general feature that we shall describe: giving
vacuum expectation values to certain baryonic operators in the UV, the theory
flows to another fixed point in the far IR. In the supergravity solution a new
``throat'' develops in the IR, at the bottom of which one generally
finds a new Sasaki-Einstein manifold\footnote{This may happen to be an orbifold of $S^5$, as will be the
case in the examples we shall discuss.}.

Following  \cite{KM},  we also propose that one may extract information about the
 one-point function (condensate)
of baryonic operators turned on in a given geometry by computing the Euclidean action of certain
instantonic D3-brane configurations in the background. In particular, we will gather evidence for the validity of this
conjecture by showing that the  exponentiated on-shell Euclidean D3-brane action quite generally reproduces
the correct scaling dimensions and baryonic charges of the baryonic operators that  acquire non-zero VEVs. 
This generalises the result of \cite{KM}, which was for the resolved conifold geometry.
Given a background geometry, one may also use these results
as a guide to predict which operators have acquired non-zero expectation values. 
We shall illustrate this for the $Y^{p,q}$ theories and their resolutions in section \ref{VEV1}. We anticipate
that a complete treatment of such instantonic D3-branes will be rather involved and subtle. In particular, one requires
a somewhat deeper understanding of the map between baryonic operators in the gauge theory and the dual objects, which
are, roughly speaking, specified by certain divisors/line bundles in the geometry. We shall
make a few more comments on this in the discussion section.

The plan of the rest of the paper is as follows.
In section \ref{sec3} we discuss generic  features
of supergravity backgrounds corresponding to baryonic branches, including some remarks on a Euclidean
D3-brane calculation that quite generally should compute baryonic condensates.
In sections \ref{toricsection} and \ref{themetrics} we provide a toric
description of Calabi-Yau metrics on various
partial resolutions recently discovered by the authors in \cite{np1}.
In section \ref{VEV1} we present the gauge
theory interpretation of the geometries previously discussed.
In section \ref{webu} we conclude and discuss briefly
some of the issues that have arisen in the paper.


\section{Baryonic branches}
\label{sec3}

\subsection{Spacetime background}
\label{spacetravel}

In this section we discuss the class of Type IIB backgrounds we
wish to consider. These will be supergravity backgrounds produced by placing
$N$ coincident D3-branes at a point on a complete asymptotically
conical Ricci-flat K\"ahler six-manifold $(X,g_X)$.
The presence of the D3-branes
induces a warp factor that is essentially the Green's function
on $(X,g_X)$; we argue that such a warp factor always exists and is unique.

The spacetime background $(M_{10},g_{10})$ we are interested in is given by the
following supersymmetric
solution of Type IIB supergravity
\bea
g_{10} & = & H^{-1/2} g_{\mathbb{R}^{1,3}}+ H^{1/2} g_X\label{metricb}\\
G_5 & = & (1+*_{10}) \diff {H^{-1}} \wedge \vol_4~.
\label{startrek}
\eea
Here $g_{\mathbb{R}^{1,3}}$ is the flat Minkowski metric, with
volume form $\vol_4$, and $(X,g_X)$ is a complete
Ricci-flat K\"ahler six-manifold. The warp factor $H$ is a
 function on $X$. If we take $H$ to be a positive constant then the
background metric (\ref{metricb}) is Ricci-flat. However, if
 we now place a stack of $N$ D3-branes parallel to
$\mathbb{R}^{1,3}$ and at the point $\mathbf{x}_0\in X$ then these
act as a source for the RR five-form flux $G_5$. The corresponding
equation of motion then gives \bea \Delta_{\mathbf{x}} H & = & -
\frac{{\cal C}}{\sqrt{\det g_X}} \, \delta^6 (\mathbf{x} -
\mathbf{x}_0)~. \label{goofy} \eea Here $\Delta$ is the Laplacian on
$(X,g_X)$, and $\mathcal{C}$ is a constant given by \bea \mathcal{C}
~= ~(2\pi)^4 g_s (\alpha')^2 N~. \label{nerd} \eea Thus
$H=G(\mathbf{x},\mathbf{x}_0)$ is a Green's function on the
Calabi-Yau $(X,g_X)$. For instance, when $X=C(Y)$ is a cone over a
Sasaki-Einstein manifold $(Y,g_Y)$ \bea g_X & =& \diff r^2 + r^2
g_Y~, \eea placing the D3-branes at the apex of the cone
$\mathbf{x}_0=\{r=0\}$ results in the following Green's
function\footnote{Since we are interested in the near-horizon
geometry, we have dropped an additive constant. Restoring this
corresponds to the full D3-brane solution.} \bea\label{Greencone}
H_{\mathrm{cone}} & = & \frac{L^4}{r^4} \eea where \bea\label{randy}
L^4 & = & \frac{\cal C}{4\vol (Y)}~. \eea This last relation is
determined by integrating $\sqrt{\det g_X} \Delta_{\mathbf{x}} H$
over the cone: the right hand side of (\ref{goofy}) gives
$-\cal{C}$, whereas the integral of the left hand side reduces to a
surface integral at infinity, which gives the relation to $\vol(Y)$.
The Type IIB solution (\ref{metricb}) is then in fact AdS$_5\times
Y$, where $L$ in (\ref{randy}) is the AdS$_5$ radius.

Assuming the Green's function $G(\mathbf{x},\mathbf{x}_0)$ on $(X,g_X)$ exists, asymptotically
it will approach the Green's function for the cone (\ref{Greencone}),
and the same reasoning as above still requires the
relation (\ref{randy}) to hold. On the other hand, the Green's function
blows up at the point $\mathbf{x}_0$. Indeed, we have
\bea
\label{explosion}
G(\mathbf{x},\mathbf{x}_0) ~= ~\frac{L_{IR}^4}{\rho(\mathbf{x},\mathbf{x}_0)^4}(1+o(1))~,
\eea
where $\rho(\mathbf{x},\mathbf{x}_0)$ is the geodesic distance from $\mathbf{x}_0$ to $\mathbf{x}$, and
\bea
L_{IR}^4~ =~ \frac{\mathcal{C}}{4\vol(S^5)}~.
\eea
The normalisation constant $L_{IR}^4$ is computed as above, noting that the
metric in a neighbourhood of $\mathbf{x}_0$ looks like flat space in polar coordinates
$\diff\rho^2+\rho^2 g_{S^5}$. If $(X,g_X)$ is only a partial resolution
of $X$ and $\mathbf{x}_0$ is a singular point, this metric is instead
$\diff\rho^2+\rho^2 g_{Z}$ where $g_Z$ is a Sasaki-Einstein metric on the link
$Z$ of the singularity. More generally one would then have\footnote{However, the general existence of the Green's
function on such a singular $(X,g_X)$ is not
guaranteed by any theorem we know of, unlike the smooth case treated below.}
$L_{IR}^4 = \mathcal{C}/4\vol(Z)$.

Due to the singular behaviour of the Green's function at the point $\mathbf{x}_0$
in (\ref{explosion}) we see that the metric (\ref{metricb}),
with $H=G(\mathbf{x},\mathbf{x}_0)$, develops an additional ``throat'' near to
$\mathbf{x}_0$, with the metric in a neighbourhood of $\mathbf{x}_0$ (with $\mathbf{x}_0$ deleted)
being asymptotically AdS$_5\times Z$. Here $Z=S^5$ if $\mathbf{x}_0$ is a smooth point.
Thus the gravity solution (\ref{metricb}) - (\ref{startrek}) has
two asymptotic AdS regions, and
may be interpreted as a renormalisation group flow from the
original theory to a new theory in the IR. 

A Green's function on a Riemannian manifold $(X,g_X)$ of dimension $n$
is by definition a function on $X\times X\setminus \mathrm{diag}(X\times X)$
satisfying:
\begin{itemize}
\item $G(\mathbf{x},\mathbf{y})=G(\mathbf{y},\mathbf{x})$, and $\Delta_{\mathbf{x}} G = 0$ for all $\mathbf{x}\neq \mathbf{y}$ with $\mathbf{y}$ fixed.
\item $G(\mathbf{x},\mathbf{y})\geq 0$.
\item As $\mathbf{x}\rightarrow \mathbf{y}$, with $\mathbf{y}$ fixed, we have
\bea
G(\mathbf{x},\mathbf{y}) ~= ~\frac{A}{\rho(\mathbf{x},\mathbf{y})^{n-2}}(1+{o}(1))
\eea
for $n=\dim_\R X>2$, where $\rho(\mathbf{x},\mathbf{y})$ denotes the geodesic distance between $\mathbf{x}$ and $\mathbf{y}$,
and $A$ is a positive constant.
\end{itemize}
Such a function doesn't necessarily always exist. However, in the
present set-up we may apply the following result of
\cite{varopoulos}: if $(X,g_X)$ is complete and has non-negative
Ricci curvature then the Green's function above exists and is finite
and bounded away from the diagonal in $X\times X$ if and only if
\bea \label{converge} \int_r^{\infty}
\frac{t}{\vol(B(t,\mathbf{y}))}~\diff t ~<~\infty \eea for all $r>0$
and all $\mathbf{y}\in X$. Here $B(t,\mathbf{y})$ is the ball of
radius $t$ and centre $\mathbf{y}$. If the volume growth of the
manifold is at least quadratic, then the integral on the left hand
side of (\ref{converge}) always converges. In our case, $(X,g_X)$ is
complete, Ricci-flat, and is asymptotically conical, which implies
the volume of any ball grows like $\rho^6$, where $\rho$ is the
distance function from any point in $X$. There is, moreover, a
unique Green's function that asymptotes to zero at infinity. The
proof of this is a simple application of the maximum principle.

The background geometries will depend on various moduli.
An asymptotically conical Ricci-flat K\"ahler metric on $X$ will generally depend on a number of
moduli. However, we note that, in contrast to the case of compact Calabi-Yau manifolds where
the moduli space is
understood extremely well, there is currently no general understanding
of the moduli space of non-compact Calabi-Yau manifolds.
In addition to the metric moduli, there are a number of flat background fields that may be turned on
without altering the solution (\ref{metricb}) - (\ref{startrek}). For instance, there is the dilaton $\phi$,
which determines the string coupling constant\footnote{Here it really is constant.}
$g_s=\exp(\phi)$. This is
paired under the $SL(2;\R)$ symmetry of Type IIB supergravity with
the axion field $C_0$. The topology of $X$ in general allows one to turn on various
topologically non-trivial flat form-fields. In particular we have the NS $B$-field, as well as the RR  two-form $C_2$
and four-form $C_4$. These play an important role in a detailed mapping between the gauge theory and geometry
moduli spaces. However, these fields will be
largely ignored in the present paper.


\subsection{Baryons and baryonic operators}
\label{baryonsgeom}

Below we recall how baryonic symmetries and baryonic particles arise
in AdS/CFT.  We also extend the proposal of \cite{KM} for the use of Euclidean
D3-branes as a means to detect non-zero expectation values of
baryonic operators in a given background geometry.

Consider a Sasaki-Einstein manifold $Y$ with
$b_3\equiv b_3(Y)=\dim H_3(Y;\R)$. By
wrapping a D3-brane on a  3-submanifold $\Sigma\subset Y$
we effectively obtain a
particle in AdS. This particle is BPS precisely when the
3-submanifold is supersymmetric, which is equivalent to the cone
$C(\Sigma)\subset C(Y)$ being a complex submanifold, or
\emph{divisor}. In
\cite{Witten:1998xy,Gukov:1998kn,Berenstein:2002ke} such wrapped
D3-branes were interpreted as \emph{baryonic particles}.
This also leads one to identify the non-anomalous
baryonic symmetries in the field theory as arising from the topology
of $Y$, as follows. Fluctuations of the RR four-form potential $C_4$
in the background AdS$_5\times Y$ may be expanded in a basis of
harmonic three-forms of $(Y,g_Y)$
 \bea
 \delta C_4 & = &
\sum_{I=1}^{b_3} \mathcal{A}_I \wedge {\cal H}_I~. \eea
Here
$\mathcal{H}_I\in\mathcal{H}^3(Y,g_Y)$ are harmonic three-forms that
are generators of the image of $H^3(Y;\Z)$ in
$\mathcal{H}^3(Y,g_Y)$.
The fluctuations give rise to $b_3$ gauge fields $\mathcal{A}_I$ in
AdS$_5$. As usual these gauge symmetries in AdS become global
symmetries in the dual field theory, and are identified precisely
with the non-anomalous baryonic symmetries $U(1)_B^{b_3}$. The
charge of a baryonic particle arising from a 3-submanifold
$\Sigma$, with respect to the $I$-th baryonic $U(1)_B$, is thus given
by \bea
 Q_I[\Sigma] ~= ~\int_\Sigma {\cal H}_I~.
\label{kubi} 
\eea

In fact, the above discussion overlooks an important  point: the D3-brane carries a
worldvolume gauge field $M$. For a D3-brane wrapping $\R_t\times\Sigma$, supersymmetry
requires this gauge field to be flat.
Thus, as originally pointed out in
\cite{Gukov:1998kn}, if $\Sigma$ has non-trivial fundamental group
one can turn on distinct flat connections on the worldvolume of the
wrapped D3-brane, and {\it a priori} each corresponds to a different baryonic
particle. These flat connections are defined on torsion line
bundles $L$ over $\Sigma$. Thus $c_1(L)\in H^2_{\mathrm{tor}}(\Sigma;\Z)$.

The dual operator that creates a baryonic particle associated to $(\Sigma, L )$
is denoted  ${\cal B}(\Sigma,L)$.
For fixed $\Sigma$ these
all have equal  baryonic charge \eqn{kubi} and
also equal R-charge, where the latter is determined by the volume
of $\Sigma$ via \cite{Berenstein:2002ke}
\bea
R(\Sigma) &= &\frac{N\pi\vol(\Sigma)}{3\vol(Y)}~.
\label{racharge}
\eea

Given a background geometry that is dual to an RG flow induced by giving expectation values
to some baryonic operators, it is natural to ask whether
it is possible to compute baryonic one-point functions
by performing some supergravity calculation.
Following the conifold example discussed in \cite{KM} we shall argue that, quite generally, a candidate for computing the VEV of
a baryonic operator is a \emph{Euclidean D3-brane} that wraps
an asymptotically conical divisor $D$ in the asymptotically conical (partial) resolution $X$,
such that $D$ has boundary $\partial D=\Sigma\subset Y$.
Indeed, taking inspiration from the Wilson loop prescription \cite{Rey:1998ik,Maldacena:1998im},
it is natural to conjecture that the holographic expectation value of a baryonic operator
 is given by the path integral of a Euclidean
  D3-brane
  with fixed boundary conditions:
\bea
\langle {\cal B} (\Sigma,L) \rangle ~ = ~ \int_{\partial D=\Sigma} {\cal D} \Psi \, \exp (-S_{D3}) ~\approx~
\exp(-S_{D3}^{\mathrm{on-shell}})  ~.
\eea

Roughly, $S_{D3}^{\mathrm{on-shell}}$ is the appropriately regularized action of
a Euclidean D3-brane, whose worldvolume $D$ has as
boundary a supersymmetric three-dimensional submanifold $\Sigma \subset Y$. 
In fact, a complete prescription for computing a baryonic condensate should take into account the
analogous extension of the torsion line bundle $L$, and thus in particular the worldvolume gauge 
field.
This is rather subtle and would take us too far afield in the present paper -- we will return to this, and related issues, in
a separate publication \cite{np3}. In the following two subsections we will show that the exponentiated
on-shell Euclidean D3-brane action obeys the following two basic properties: (1)
it reproduces the correct scaling dimension, and (2) it carries the correct baryonic charges. 
In the computation of the scaling dimension we will \emph{formally} set the worldvolume gauge 
field to zero, in line with the comment above. One might worry\footnote{We are grateful to the referee for suggesting that we emphasize this issue.} that in general the gauge field contributes a divergent term to the large radius 
expansions we discuss below. However, since the 
result with zero gauge field already produces the expected scaling dimension of the dual operator, 
it is natural to conjecture that including the worldvolume gauge field does not alter this result.
This will be shown in detail in the paper \cite{np3}.


\subsection{Scaling dimensions of baryonic condensates}
\label{greeny}

The real part of the  Euclidean D3-brane action is given by the
Born-Infeld term \bea \label{barry} S_{BI} = T_3\int_D \diff^4\sigma
\sqrt{\det(h+M)}~.
 \eea
Here $D$ is the D3-brane worldvolume, with local coordinates
$\sigma_{\alpha}$, $\alpha=1,\ldots,4$, and supersymmetry requires
$D$  to be a divisor in $X$. $T_3$ is the D3-brane tension, given by
\bea T_3 ~= ~\frac{1}{(2\pi)^3\alpha'^2g_s}~. \label{tension} \eea
$h$ is the first fundamental form {\it i.e.} the induced metric on
$D$ from its embedding into spacetime $\iota:D\hookrightarrow
(M_{10},g_{10})$. $M$ is the worldvolume gauge field that we will formally
set to zero.
Then the real part of the action reduces to 
\bea
 S_{BI} & = & T_3\int_D
 \diff^4\sigma \sqrt{\det g_D} H
 \label{cuteaction}
 \eea
where $g_D$ is the metric induced from the embedding of $D$ into $(X,g_X)$.
Below we  show that
the integral in (\ref{cuteaction}) is always divergent and thus
needs to be regularised\footnote{See \cite{Karch:2005ms} for a careful treatment of 
holographic renormalisation of probe D-branes in AdS/CFT.}. 
We evaluate the integral up to a large UV cut-off
$r=r_c$. This will show that the action has precisely the
divergence, near infinity $r_c\rightarrow\infty$, expected for a baryonic operator
that has acquired a non-zero expectation value. As mentioned at the end of 
section \ref{baryonsgeom}, this calculation of the scaling dimension is rather formal since 
we have set the worldvolume 
gauge field $M$ to zero. A complete treatment that also includes the gauge field will appear in \cite{np3}.
Our analysis below will also lead to a simple \emph{necessary} condition
for the holographic condensate to be non-vanishing.

At large $r$, the geometry is asymptotically
AdS$_5\times Y$, where $r$ becomes a radial coordinate in AdS$_5$. Then,
following\footnote{Strictly speaking,
the prescription of \cite{KW2}, which is an extension of the original prescriptions
of \cite{Gubser:1998bc,Witten:1998qj}, is formulated for
supergravity modes. Here we shall assume that this remains valid for the
intrinsically stringy field describing a (Euclidean) D3-brane, as in \cite{KM,Benna:2006ib}.}
\cite{KW2}, one can
interpret the asymptotic coefficients in the expansion of a field $\Phi$ near the AdS$_5$ boundary
\bea
\Phi~ \sim ~\Phi_0 \, r^{\Delta- 4} + A_\Phi \, r^{-\Delta}~,
\label{drwho}
\eea
as corresponding to
the source  of a dual operator ${\cal O}_\Delta$ and its one-point function, respectively.
Here $\Delta$ is the scaling dimension of  ${\cal O}_\Delta$.
In particular, if $\Phi_0$ vanishes, the background is dual to an RG flow
triggered purely by the
condensation  of the operator  ${\cal O}_\Delta$, without explicit insertion of the
operator into the UV Lagrangian.

Let $D[r_c]$ denote the compact manifold with boundary defined by
cutting off a divisor $D$ at some large radius $r_c$. We then define
\bea S[r_c,\mathbf{x}_0] & = & T_3 \int_{D[r_c]} \diff^4 \sigma  \sqrt{\det
g_D} \, G(\mathbf{x},\mathbf{x}_0)~, \eea
where we regard this as depending on
the position of the stack of D3-branes $\mathbf{x}_0\in X$. We then show that the
following result is generally true: 
\be
\label{pinocchio}
\exp(-S[r_c,\mathbf{x}_0]) ~=~ \left\{
\begin{array}{ll}
  0 & \mathrm{if}\quad   \mathbf{x}_0\in D\\[2mm]
  \mathcal{O}\left(r_c^{-\Delta(\Sigma)}\right) &  \mathrm{if}\quad  \mathbf{x}_0 \notin D~.
\end{array}
\right. \ee
Here
\bea
\label{conformaldimension}
\Delta(\Sigma) & = &
\frac{N\pi\vol(\Sigma)}{2\vol(Y)}
\eea
is the conformal dimension of
any baryonic operator associated to $\Sigma$, under the correspondence
discussed in section \ref{baryonsgeom}. In particular, this result is
insensitive to the choice of torsion line bundle $L$ on $\Sigma$.
It is interesting to note that if we keep the divisor $D$ fixed and 
regard $\exp(-S[r_c,\mathbf{x}_0])$ as depending on the position 
of the  D3-branes $\mathbf{x}_0$, then from  (\ref{pinocchio}) we may deduce that this has a zero along $D$, and 
is otherwise non-singular.  These properties are compatible with the interpretation that a baryonic 
condensate is in fact a section of the divisor bundle ${\cal O}(D)$.

The proof of (\ref{pinocchio}) is rather simple.
Suppose first that $\mathbf{x}_0\in D$. In a small ball around a smooth
point
$\mathbf{x}_0$ in $X$ the Green's function behaves as
\bea
H ~= ~\frac{L^4_{IR}}{\rho^4}(1+o(1))\qquad\quad  L^4_{IR} ~= ~\frac{\mathcal{C}}{4\vol(S^5)}
\eea
where $\rho$ is the geodesic distance from $\mathbf{x}_0$.
A neighbourhood of $\mathbf{x}_0$ in $D$ looks like
$\R^4$ with radial coordinate $\rho\mid_D$.
Let us evaluate the integral in a compact annular domain $V(\epsilon)$, defined
by $0<\epsilon\leq \rho\mid_D\leq\delta$. Here we shall hold $\delta$ small and fixed, and
examine the integral in the limit $\epsilon\rightarrow0$:
\bea
\int_{V(\epsilon)} \!\!\!\!\!\diff^4 \sigma \sqrt{\det g} G(\mathbf{x},\mathbf{x}_0)  =  \int_{V(\epsilon)}
\!\!\!\!\frac{L^4_{IR}}{\rho^4}\rho^3 (1+o(1))\diff \rho \ \diff\vol_{S^3}
\sim  L^4_{IR} \vol (S^3)
\log (1/\epsilon)~.
\label{bigballs}
\eea
Since the Green's function is positive everywhere, this logarithmic
divergence at $\epsilon=0$ (that is at $\mathbf{x}=\mathbf{x}_0)$ cannot be cancelled,
and we have proved the first part of (\ref{pinocchio}).

Suppose now that $\mathbf{x}_0\notin D$. Then the Green's function $G(\mathbf{x},\mathbf{x}_0)$ is
positive and bounded everywhere on $D$. Let us cut the integral in
two. We integrate first up to $r_0<r_c$, where $r_0$ will be held
large and fixed, and then integrate from $r_0$ to $r_c$. Let the latter
domain be denoted $V(r_0,r_c)$.
The integral up to $r_0$ is finite. The integral over $V(r_0,r_c)$ is, in the
limit $r_c\rightarrow\infty$,
\bea
\int_{V(r_0,r_c)} \diff^4 \sigma \sqrt{\det g} G(\mathbf{x},\mathbf{x}_0)
~\sim~ \int_{r_0}^{r_c} \frac{L^4}{r^4}r^3\diff r\vol(\Sigma) ~\sim ~L^4 \vol(\Sigma)\log r_c~.
\eea
Now recalling the normalisation (\ref{nerd}) and
(\ref{tension}) that we gave earlier,
we compute
\bea
T_3 \mathcal{C} = 2\pi N~.
\eea
Inserting this into (\ref{randy}), we arrive at
\bea
S[r_c,\mathbf{x}_0]\sim T_3 L^4 \vol(\Sigma)\log r_c &=&
\Delta (\Sigma) \,\log r_c~,
\eea
showing that indeed
\bea
\exp(-S[r_c,\mathbf{x}_0]) \sim  A \, r_c^{-\Delta(\Sigma)}~
\eea
gives the leading behaviour as $r_c\rightarrow\infty$.
We interpret this result as a signal that a baryonic operator ${\cal B}(\Sigma,L)$ of conformal dimension
$\Delta$ has acquired a vacuum expectation value $\langle{\cal B}(\Sigma,L) \rangle \propto A$.
When $\mathbf{x}_0\in D$ the above analysis shows that $A=0$ identically and thus the
condensate certainly vanishes. Thus $\mathbf{x}_0\notin D$ is a \emph{necessary}
condition for non-vanishing of the condensate.


\subsection{Baryonic charges of baryonic condensates}
\label{imagine}

We will now consider the Chern-Simons part of the Euclidean D3-brane
action, which upon setting $M=0$, reduces to
 \bea
\label{genCSaction} S_{CS} & = & i\mu_3\int_D C_4 ~. \eea Here
$C_4$ is the RR potential and the D3-brane charge is given
by\footnote{See,
for example, \cite{johnson}.}
\bea
\mu_3 ~=~\frac{1}{(2\pi)^3\alpha'^2}~.
\eea
A careful analysis of the remaining
terms, involving $C_2$ and $C_0$ RR potentials, will be presented
elsewhere \cite{np3}.

Given that our background geometries are non-compact, it is important to consider the
role of the boundary conditions for the background fields.
Asymptotically we approach an AdS$_5\times Y$
geometry. This 
describes the superconformal theory that
is being perturbed, and in particular the boundary values of fields
on $Y$ specify this superconformal theory.
We  thus require all background fields to approach well-defined fields on
AdS$_5\times Y$ at infinity.
To make this statement more precise, we may cut off the asymptotical conical
geometry at some large radius $r_c$; the boundary is
denoted $Y_{r_c}$, which for large $r_c$ is diffeomorphic (by not isometric in general) to the
Sasaki-Einstein boundary $(Y,g_Y)$.
The restriction of all fields to $Y_{r_c}$, or rather $\R^{1,3}\times Y_{r_c}$,
 should then give well-defined smooth
fields on $Y$ in the limit $r_c\rightarrow \infty$, and these values specify the
superconformal theory in this asymptotic region.

Note that for the conical geometry $C(Y)$, which corresponds to the AdS$_5\times Y$ background itself,
the internal RR flux is proportional to the volume form on $(Y,g_Y)$.
Thus, in particular, there is no globally defined $C_4$ such that $\diff C_4=G_5$.
Since $G_5^{\mathrm{cone}}\mid_X=\vol_Y$, a natural gauge choice is to
take $C_4$ (locally) to be a pull-back from $Y$ under the projection
$\pi:C(Y)\cong \R_+\times Y\rightarrow Y$.
By picking a trivialisation over local patches $U\subset Y$,
the integral of the corresponding $C_4^\mathrm{cone}$
over $D\cap \pi^{-1}(U)$ vanishes, since $D$ is a cone and the
contraction of $\partial/\partial r$ into $C_4^\mathrm{cone}$ is zero by construction.
For a general asymptotically conical background $(X,g_X)$ with the $N$ D3-branes
at the point $\mathbf{x}_0\in X$, the corresponding $G_5$ will approach asymptotically the conical
value. Thus we may choose a gauge $C_4^{\mathrm{background}}$ which
approaches the above gauge choice for $C_4^{\mathrm{cone}}$ near infinity.
With this gauge choice we deduce that the integral
\bea\label{number}
i\mu_3\int_D C_4^{\mathrm{background}}
\eea
is finite\footnote{When $X$ is toric, using
symplectic coordinates one can show that there is a gauge in which $C_4$
has vanishing pull-back to any asymptotically conical
toric divisor. In particular, we may locally write
 $C_4=\diff \phi_1\wedge \diff \phi_2 \wedge \diff \phi_3 \wedge
A$ for some one-form $A$.}.

In general, to any background choice of $C_4^{\mathrm{background}}$ we may add a
closed four-form. If this four-form is not exact one obtains a physically distinct background.
Indeed, recall that the basic gauge transformation of $C_4$ is the shift
\bea
\label{C4shift}
C_4 ~\rightarrow ~C_4 + \diff K
\eea
where $K$ is a three-form. The above integral (\ref{number}) then clearly depends on the choice of
gauge, since
\bea
i \mu_3\int_D C_4 ~\to ~ i \mu_3\int_D C_4~ +~ i \mu_3\int_{\Sigma} K_Y~,
\label{intKY}
\eea
where $K_Y=K\mid_Y$ is the restriction to $Y$ of the three-form
$K\in\Omega^3(X)$. As discussed above,
we should consider only those
gauge choices for $C_4$ that give a well-defined form on $Y$, implying
that $\diff K$ also restricts to a well-defined form on $Y$.
We may thus take $K$ itself to be
well-defined in the limit $Y=\lim_{r_c\rightarrow\infty} Y_{r_c}$, modulo an exact part that has no such restriction.
The exact part may diverge in the limit, but at the same time it
drops out of the integral (\ref{intKY}) since $\Sigma$ is compact, where
more precisely we should define the integral as the limit of an integral over
$\Sigma_{r_c}$. Note that the phase shift (\ref{intKY}) depends only on $K_Y$, and not on the
extension $K$ of $K_Y$ over $X$.

On the other hand, true \emph{symmetries} are gauge transformations that do not change the fields
at infinity. Thus we should consider a \emph{fixed} gauge choice for $C_4\mid_Y$ on the
AdS boundary, and gauge transformations whose generator $K\in \Omega^3(X)$ is such that $\diff K_Y=0$.
Gauge transformations of $C_4$  whose generators $K$
vanish at infinity act trivially on physical states.
Thus  shifts (\ref{C4shift}) where $K\mid_{Y}=0$ produce physically equivalent $C_4$ fields.
Indeed, recall that global symmetries in gauge theories arise from gauge symmetries whose generators do
not vanish at infinity but that leave the fields fixed at infinity\footnote{Notice that this discussion
parallels a similar discussion in \cite{Atiyah:2001qf}.}.
We therefore identify these transformations of the background $C_4$ as
the \emph{non-anomalous baryonic symmetries} in the gauge theory.

We may pick a natural representative for a class in $H^3(Y;\R)$ using
the Hodge decomposition
\bea
\Omega^3(Y)~=~\diff\Omega^2(Y)\oplus \mathcal{H}^3(Y,g_Y)\oplus\delta\Omega^4(Y)\eea
on $(Y,g_Y)$. We may then write any closed $K_Y$ uniquely as
\bea
K_Y ~= ~K_Y^{\mathrm{harm}} + \diff \lambda
\eea
where $K_Y^{\mathrm{harm}}\in \mathcal{H}^3(Y,g_Y)\cong H^3(Y;\R)$.
Of course, $\int_{\Sigma}\diff \lambda=0$. Thus, although there is an infinite set of background gauge-equivalent $C_4$ fields on $X$
that approach a given boundary gauge choice on $Y$, the
integral of $C_4$ over any $D$ depends only on
a finite dimensional part of this space, namely the harmonic part of $K_Y$.
We may then expand
\bea
\mu_3 \, K_Y^{\mathrm{harm}} & = &\sum_{I=1}^{b_3} \beta^I \, {\cal H}_I
\eea
where recall that ${\cal H}_I \in \mathcal{H}^3(Y,g_Y)$ form an integral basis for the
image $H^3(Y;\Z)\rightarrow H^3(Y;\R)$.
Notice that shifting the periods of $C_4$ by an integer multiple of
$(4\pi^2\alpha')^2$ (large gauge transformations) does not change the quantum measure
$\exp(-S)$.
Thus the global symmetry group arising from gauge symmetries of $C_4$ is, more precisely,
the compact abelian group $H^3(Y;U(1))\cong U(1)^{b_3}$, and in particular the $\beta^I$ are periodic
coordinates.
Notice that these harmonic three-forms are the same as those
appearing in the KK ansatz discussed in section \ref{baryonsgeom}, that give rise
to the baryonic symmetries. We conclude that the effect of the above gauge transformation is to shift
\bea
\exp \left[ i \mu_3 \int_D C_4 \right]  ~\to  ~
\exp \left(  i \beta^I Q_I[\Sigma]\right) \, \exp \left[ i \mu_3 \int_D C_4 \right] ~.
\eea
This is a global transformation in the boundary SCFT, where
$Q_I[\Sigma]$  is the baryonic charge of the baryonic operator ${\cal B}(\Sigma,L)$
with respect to the $I-$th baryonic  $U(1)_B$ \cite{tilings}.


\section{Toric description of $Y^{p,q}$ partial resolutions}
\label{toricsection}

So far our discussion has been rather general. In the remainder of
the paper we discuss a set of examples, namely the $Y^{p,q}$ theories.
In the present section
we review the toric geometry of $Y^{p,q}$ \cite{toricpaper} and discuss several classes of
(partial) resolutions of the corresponding isolated Gorenstein singularities.
We present explicit asymptotically conical Ricci-flat K\"ahler
metrics on these partial resolutions in section \ref{themetrics}.
The results of section \ref{greeny} concerning the vanishing
of certain baryonic condensates due to the behaviour of the
Green's function in fact translate into simple
pictures in toric geometry. For the $Y^{p,q}$ theories, the map from
toric conical divisors $D=C(\Sigma)$, with link $\Sigma$ equipped with a torsion
line bundle $L$,
to a class of baryonic operators constructed simply as determinants of
the bifundamental fields
is known from the original papers
\cite{quiverpaper, tilings}. The toric pictures for
the partial resolutions referred to above then immediately
allow one to deduce which bifundamental fields do \emph{not} obtain a VEV
for that background. In the examples we discuss this simple sufficient condition
for the condensate to vanish thus leads to predictions that may easily be
checked directly in the quiver gauge theory. In section \ref{VEV1} we verify
these predictions by giving VEVs to the relevant bifundamental fields,
and determining where the resulting theory flows in the far IR. The results
agree precisely with the geometry of the partial resolutions.

\subsection{Toric geometry and the $Y^{p,q}$ singularities}\label{Ypq}

We begin by briefly reviewing the geometry of toric Gorenstein
(Calabi-Yau) singularities, focusing in particular on the $Y^{p,q}$
geometries and their toric resolutions.

\begin{figure}[ht!]
  \epsfxsize = 4cm
  \centerline{\epsfbox{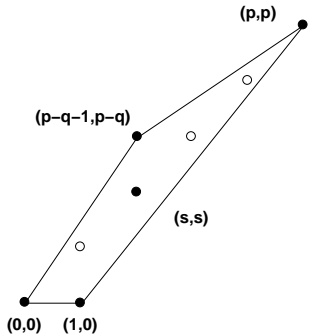}}
  \caption{Toric diagram of a $Y^{p,q}$ singularity, with internal point $(s,s)$ shown.
Here $0<s<p$.}
  \label{troia}
\end{figure}

A toric Gorenstein
singularity in complex dimension three is specified by a convex
lattice polytope $\Delta\subset \R^2$. Such a polytope may be
specified by a set of vectors $w_a\in\Z^2\subset\R^2$,
$a=1,\ldots,d$, which are the defining external vertices of the
polytope. More precisely, there is a 1-1 correspondence between such
singularities and $SL(2;\Z)$ equivalence classes of convex lattice
polytopes, where the origin may be placed anywhere in the lattice.
Here $SL(2;\Z)$ acts on $\Z^2\subset\R^2$ in the obvious way. A
choice of lattice polytope for the $Y^{p,q}$ singularities is shown
in Figure \ref{troia}.
The external points of the lattice polytope are, moving anti-clockwise starting from
the lower right corner, given by:
$w_1=(1,0)$, $w_2=(p,p)$, $w_3=(p-q-1,p-q)$,  $w_4=(0,0)$. Thus $d=4$ for
the $Y^{p,q}$ singularities.

The geometry is recovered from the lattice polytope by a form of Delzant's construction.
One first defines the three-vectors $v_a=(1,w_a)\in\Z^3$. These define a linear map
\bea
A:&& \mathbb{R}^d\rightarrow \mathbb{R}^3\nonumber\\
&& e_a\mapsto v_a\eea
where $\{e_a\}_{a=1,\ldots,d}$ denotes the standard orthonormal basis of $\mathbb{R}^d$.
Let $\Lambda\subset\mathbb{Z}^3$
denote the lattice spanned by the $\{v_a\}$ over $\mathbb{Z}$. This is of maximal rank,
since the polytope $\Delta$ is convex. There is then an induced map of tori
\bea
U(1)^d\cong \mathbb{R}^d/2\pi\mathbb{Z}^d\longrightarrow \mathbb{R}^3/2\pi\mathbb{Z}^3\cong U(1)^3\eea
where the kernel is a compact abelian group
$\mathcal{A}$, with $\pi_0(\mathcal{A})\cong \Gamma\cong \mathbb{Z}^3/\Lambda$.

Using this data we may construct the geometry as a K\"ahler
quotient. Thus, using the flat metric, or equivalently standard
symplectic form, on $\C^d$, we may realise \bea\label{quotient} C(Y)
~= ~\mathbb{C}^d \ //_0 \ \mathcal{A}~.\eea Here $\mathcal{A}\cong
U(1)^{d-3}\times\Gamma\subset U(1)^d$ acts holomorphically and
Hamiltonianly on $\mathbb{C}^d$. The subscript zero in
(\ref{quotient}) indicates that we take the K\"ahler quotient at
level zero. The origin of $\C^d$ projects to the singular point in
$C(Y)$, and the induced K\"ahler metric on $C(Y)\cong\R_+\times Y$
is a cone. Moreover, the quotient torus $U(1)^d/\mathcal{A}\cong
U(1)^3$ acts holomorphically and Hamiltonianly on this cone, with
moment map \bea
\mu &:& C(Y)\rightarrow\mathtt{t}^*\cong\R^3\\
&& \mu(C(Y))~=~\mathcal{C}^*~.
\eea
Here $\mathtt{t}^*\cong \R^3$ denotes the dual Lie algebra for $U(1)^3$.
The image of the moment map $\mathcal{C}^*\subset\R^3$ is a convex rational polyhedral cone
formed by intersecting $d$ planes through the origin of $\R^3$.
These bounding planes (or \emph{facets}) of the cone have inward pointing normal vectors
precisely the set $\{v_a\}$, and we have thus come full circle.

The quotient (\ref{quotient}) may be written explicitly in GLSM terms as follows. One computes
a primitive basis for the kernel of $A$ over $\Z$ by finding all
solutions to
\bea
\sum_a Q_{I}^a  v_a = 0\eea
with $Q_{I}^a\in \Z$, and such that for each $I$ the $\{Q_I^a\mid a=1,\ldots,d\}$ have no
common factor. The number of solutions, which are indexed by $I$, is
$d-3$ since $A$ is surjective; this latter fact again
follows from convexity. One then has
\bea\label{reduction}
X ~=~ \mathcal{K}_{\xi}/\mathcal{A}
~\equiv~ \C^d \ //_{\xi} \ \mathcal{A}\eea
with
\bea
\mathcal{K}_{\xi} \equiv \left\{(z_1,\ldots,z_d)\in\C^d\mid\sum_{a}Q_{I}^a |z_a|^2 = \xi_I\right\}\subset \C^d\eea
where $z_a$ denote standard complex coordinates on $\C^d$ and the charge matrix
$Q_{I}^a$ specifies the torus embedding $U(1)^{d-3}\subset U(1)^d$. In GLSM
terms, the matrix $Q_I^a$ is the charge matrix, and the set $\mathcal{K}_{\xi}$ is
the space of solutions to the D-term equations. The cone $C(Y)$ is given by setting
$\xi=(\xi_1,\ldots,\xi_{d-3})=0$.

By instead taking the K\"ahler quotient (\ref{quotient}) at level
$\xi\neq 0$ we obtain various (partial) resolutions of the
singularity $C(Y)$. In fact, to \emph{fully} resolve the singularity
we must enlarge the above K\"ahler quotient to include \emph{all}
lattice points $w_{\alpha}\subset \Delta\cap\Z^2$, $\alpha=1,\ldots,D$,
rather than simply the external vertices $w_a$. We then follow
precisely the same procedure as above, to obtain a K\"ahler quotient
of $\C^D$ with $D-3$ FI parameters. Here $D=d+\mathcal{I}$, where
$\mathcal{I}$ is the number of internal points of the toric diagram.
For example, for $Y^{p,q}$ this number is $\mathcal{I}=p-1$. It is
not too difficult to show that $d=3+b_3(Y)$ and
$\mathcal{I}=b_4(X)$, where $X$ is any complete toric resolution of
the singularity. In this larger K\"ahler quotient the image
$\mathcal{C}^*$ of $X$ under the moment map is more generally a
rational convex polyhedron. The bounding planes are precisely the
images of the \emph{toric divisors} in $X$ -- that is, the divisors
that are invariant under the $U(1)^3$ action. These are divided into
non-compact and compact, which number $d$ and $\mathcal{I}$,
respectively. By considering a strict subset of the set of all
lattice points in $\Delta$ we obtain only partial resolutions by
taking the moment map level $\xi\neq 0$. However we choose to
present the singularity as a K\"ahler quotient, the space of FI
parameters (moment map levels) that lead to non-empty quotients form
a convex cone, subdivided into conical chambers $\{C\}$. Passing
from one chamber into another across a wall, the quotient space
undergoes a small birational transformation. We shall see some
examples of this momentarily.

It is rather well-known that the chambers correspond to different
\emph{triangulations} of the toric diagram $\Delta$. The graph
theory dual of such a subdivision of the toric diagram is called the
pq-web in the physics literature. That is, one replaces faces by
vertices, lines by orthogonal lines, and vertices by faces. The
corresponding subdivision of $\R^2$ into convex subsets is in fact
precisely the projection of the image of the moment map
$\mathcal{C}^*\subset\R^3$ onto $\R^2$. One can do this canonically
here precisely due to the Calabi-Yau condition, which singles out
the vector $(1,0,0)$ one uses to construct the projection. Thus the
pq-web is a literal presentation of the Calabi-Yau: the moment map
image $\mathcal{C}^*$, which in general is a non-compact convex
polyhedron in $\R^3$, describes the quotient by the torus action
$U(1)^3$, and the pq-web is a projection of this onto $\R^2$. In
particular, the bounding planes of $\mathcal{C}^*$, which recall are
the images of the toric divisors, map onto the convex polytopic
regions in the pq-web. This allows one to map complicated changes of
topology into simple pictures that may be drawn in the plane. This
is why toric geometry is so useful.

\begin{figure}[ht!]
  \epsfxsize = 12cm
  \centerline{\epsfbox{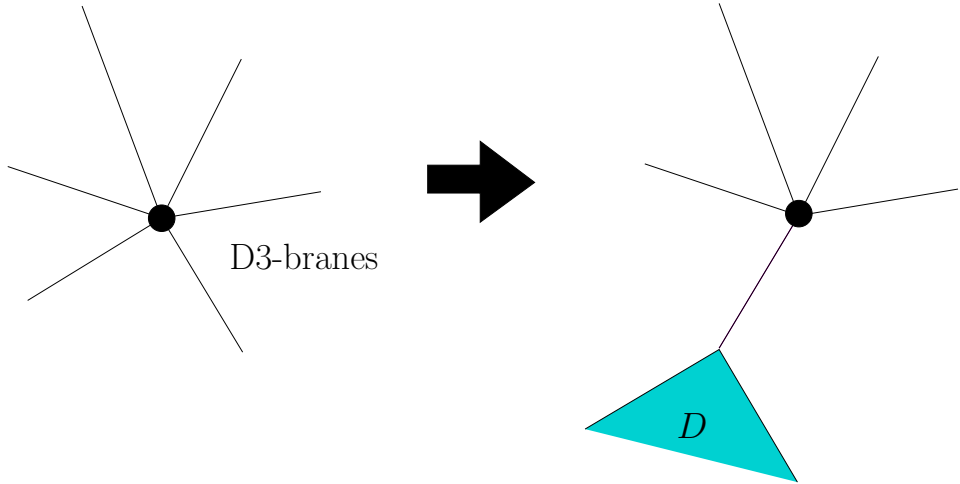}}
  \caption{On the left: pq-web with D3-branes at a toric singularity.
   On the right: a partially resolved geometry,
  with D3-branes localised at a residual singularity.
  If a toric divisor $D$ asymptotic to $C(\Sigma)$ intersects the point-like
  D3-branes, the corresponding baryonic operators do not acquire a VEV. On the other hand,
toric divisors $D$ that do not intersect the D3-branes may give rise to a condensate, as denoted
by the shaded region.}
  \label{piggy}
\end{figure}

Assuming there is an asymptotically conical Ricci-flat K\"ahler metric
for a given (partial) toric resolution $X$, we may then use
the pq-web to  give a pictorial representation of the corresponding flow geometry.
A choice of point $\mathbf{x}_0\in X$ where the $N$ D3-branes are placed determines
a choice of point\footnote{Note that vertices in the pq-web really are points
in $X$, but that points on a line in the pq-web are images of circles in $X$,
points on an open face are images of two-tori in $X$, and points ``out of the page''
(recall the pq-web is a projection of $\mathcal{C}^*$)
are images of three-tori.} in the pq-web.
Thus, using the results of section \ref{greeny}, one obtains a simple
pictorial representation of which toric divisors lead to zero condensates --
see Figure \ref{piggy}.

We decorate the pq-web with a blob, representing the location of the point-like
stack of D3-branes, and shade the divisors that do not intersect the latter.
Notice that when the D3-branes are at the conical singularity it is clear
from the picture that
no operators may have a VEV -- all toric divisors intersect the origin and thus
must have zero condensate. This is as one expects of course, since the diagram on
the left of Figure \ref{piggy} corresponds to the superconformal field theory.
Note also that the existence argument for the Green's function
presented in section \ref{spacetravel} applied only to \emph{smooth} $X$.
When $X$ is singular, as in Figure \ref{piggy}, we do not know of any general theorems.
However, at least for partial resolutions that contain at worst orbifold
singularities, the theorems referred to in section \ref{spacetravel}
presumably go through without much modification. For the $Y^{p,q}$
partial resolutions we shall restrict our attention to, we shall indeed encounter
at worst orbifold singularities.

\subsection{Small partial resolutions}\label{small}

\begin{figure}[!ht]
\vspace{5mm}
\begin{center}
\epsfig{file=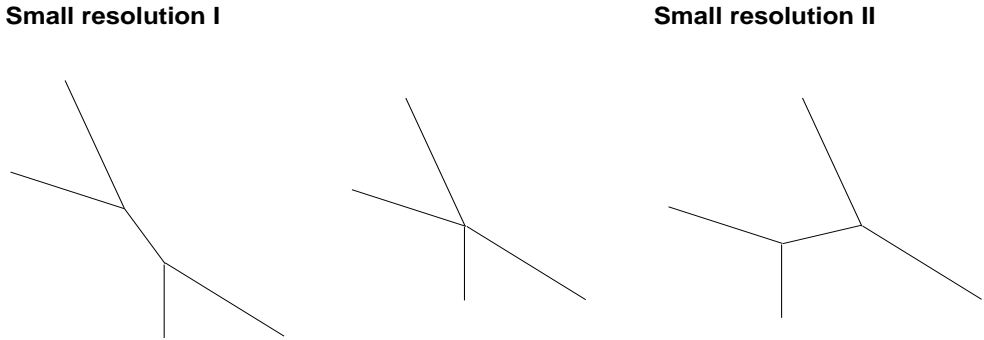,width=13cm,height=4.5cm}
\end{center}
\caption{The pq-webs for the cone $C(Y^{p,q})$ and its
 two small partial resolutions.}
\vspace{5mm}
\label{iowa}
\end{figure}

In the following two subsections we examine a simple set of partial
resolutions of the $Y^{p,q}$ singularities, starting with the
partial resolutions that correspond to the minimal presentation of
the singularity \cite{toricpaper}. Thus we realise $C(Y^{p,q})$ as a
K\"ahler quotient $\C^4\ //_0 \ U(1)$. Explicitly, the charge vector
is $Q=(p,-p+q,p,-p-q)$, with the corresponding D-term equation
given by \bea \mathcal{K}_{\xi} =
\{p|z_1|^2-(p-q)|z_2|^2+p|z_3|^2-(p+q)|z_4|^2 & = & \xi\}~.
\label{dterm} \eea The convex cone of FI parameters is the real line
$\R$, parameterised by $\xi$, and this is separated into two
chambers $C_I=\{\xi>0\}$ and $C_{II}=\{\xi<0\}$. The pq-webs are
shown in Figure \ref{iowa}.

Setting $\xi=0$ gives the K\"ahler cone,
whose corresponding Ricci-flat K\"ahler
 cone metric was constructed explicitly in \cite{paper2}. The two partial
resolutions corresponding to the two chambers will be referred to
as small partial resolution $I$, $II$, respectively. In \cite{np1}
we constructed explicit asymptotically conical Ricci-flat K\"ahler metrics
on these partial resolutions. However, the construction did not use toric
geometry. Thus, in the following two subsections we describe more explicitly
the geometry of each partial resolution in order to make contact with the
metrics of \cite{np1}, which will subsequently be presented in section \ref{themetrics}.

\subsubsection{Small partial resolution I}
\label{smally1}

Let us first consider $\xi>0$. In this case we partially resolve
the conical singularity by blowing up a $\mathbb{CP}^1=S^2$.
Explicity, this exceptional set is cut out by $\{z_2=z_4=0\}\subset\C^4$.
The D-term in (\ref{dterm}), modulo the $U(1)$ action, then clearly
gives a copy of $\mathbb{CP}^1$, with size determined by $\xi$. In fact, the whole space $X$
is  a holomorphic
$\C^2/\Z_p$ fibration over $\mathbb{CP}^1$, where $\Z_p\subset U(1)\subset SU(2)
\curvearrowright\C^2$.
One can deduce this explicitly
from the K\"ahler quotient construction, much as in \cite{toricpaper}.
An explicit Ricci-flat K\"ahler metric, that is asymptotic to
the conical metric over $Y^{p,q}$, was constructed on this bundle
in \cite{np1} -- we shall give the metric in section \ref{themetrics}.
The precise fibration structure was also
spelled out explicitly in reference \cite{np1}.

The pq-web is drawn
on the left hand side of Figure \ref{iowa}.
 The line segment in this picture is the image of the exceptional $\mathbb{CP}^1$
at $z_2=z_4=0$, and has length given roughly by $\xi$. The ends of this line segment
are two vertices corresponding to the north and south poles of $\mathbb{CP}^1$, which
is acted on isometrically by $SU(2)$. Since the whole $\mathbb{CP}^1$ is a
locus of orbifold singularities,
these two vertices are singular points, with tangent cones being $\C\times\C^2/\Z_p$.
This follows from the above fibration structure, but one may also
deduce this straightforwardly from the toric diagram by applying
Delzant's construction. This realises a neighbourhood
of either point as the quotient $\C^3/\Z_p=\C\times\C^2/\Z_p$, as the
reader may easliy verify. This is precisely the local geometry
of an $\mathcal{N}=2$ $A_{p-1}$ singularity.

\subsubsection{Small partial resolution II}
\label{smally2}

Next we consider $\xi<0$. Here one instead blows up an exceptional
weighted projective space, cut out by $\{z_1=z_3=0\}\subset\C^4$.
The details, however, depend on the parity of $p+q$.

Suppose first that $p+q$ is odd. In this case the $U(1)$ action
on $\{z_a\in \mathcal{K}_{\xi}\mid z_1=z_3=0\}$ is effective, and we obtain the weighted projective
space $\mathbb{WCP}^1_{[p-q,p+q]}$ as exceptional set.
The partial resolution is then a certain holomorphic $\C^2$
orbifold fibration over this. Precisely, this is given by
\bea
K_{\mathbb{WCP}^1_{[p-q,p+q]}}^{1/2}\times_{U(1)} \C^2\eea
where $K_M$ generally denotes the canonical orbifold line bundle of $M$,
and the $U(1)$ action on $\C^2$ above is the diagonal $U(1)\subset U(2)
\curvearrowright\C^2$.
No Ricci-flat K\"ahler metric is known on this space in general.

Suppose instead that  $p+q$ is even.
In this case the $U(1)$ action on
$\{z_a\in \mathcal{K}_{\xi}\mid z_1=z_3=0\}$ is not effective, but rather factors through a
$\Z_2$ subgroup. This means that $\{z_a\in \mathcal{K}_{\xi}\mid z_1=z_3=0\}/U(1)$ is the
weighted projective space
$\mathbb{WCP}^1_{[(p-q)/2,(p+q)/2]}$. The partial resolution is
then a holomorphic $\C^2/\Z_2$ fibration over this, given by
\bea
K_{\mathbb{WCP}^1_{[(p-q)/2,(p+q)/2]}}\times_{U(1)} \C^2/\Z_2~.\eea
Here the $U(1)$ acts effectively and diagonally on $\C^2/\Z_2$.
In particular, the fibre over a generic (non-singular) point
is now $\C^2/\Z_2$, which is the $A_1$ singularity, rather
than $\C^2$. An explicit Ricci-flat K\"ahler metric was
constructed on this orbifold in \cite{np1} and will be
reviewed in the next section.

The pq-web is given on the right hand side of Figure \ref{iowa}.
 The line segment corresponds
to the weighted projective space exceptional set (or zero-section in
the above fibration description), with length roughly
given by $\xi$. The two vertices correspond to the two singular
points of the weighted projective space. The local geometry
around these points may be determined either via the above
orbifold fibration structure, or directly via the toric diagram.
The latter may be more palatable.
Let us
consider the singular point on the weighted projective space with
local orbifold group $\C/\Z_{p-q}$ or $\C/\Z_{(p-q)/2}$  (the other point will be similar).
In either case the toric diagram describing the local geometry
is given by the triangular lattice polytope with
external vertices $w_4=(0,0)$, $w_3=(p-q-1,p-q)$, $w_1=(1,1,0)$ -- see Figure \ref{Ypic}.
The kernel of the corresponding map of tori, which is a finite subgroup, is generated by
the vector $[-2/(p-q),1/(p-q),1/(p-q)]$. Thus the local geometry
is $\C^3/\Z_{p-q}$, where $\Z_{p-q}\subset U(1)$ is embedded inside
$SU(3)$ with weights $(-2,1,1)$.
Note that, independently of the parity of $p-q$, the fibre over the singular point on
the exceptional set is $\C^2/\Z_{p-q}$,
with $\Z_{p-q}\subset U(1)$ embedded into the diagonal of $U(2)$.
This may be seen explicitly from the above orbifold fibration also --
for details, see \cite{np1}.

Thus the two points have local geometry $\C^3/\Z_{p-q}$,
$\C^3/\Z_{p+q}$, where the two abelian subgroups are embedded inside
$U(1)\subset SU(3)\curvearrowright\C^3$ with weights $(-2,1,1)$. Note that these
are both $\mathcal{N}=1$ orbifolds, rather than the $\mathcal{N}=2$
orbifolds obtained for $\xi>0$ in the previous subsection.


\subsection{Canonical partial resolutions}
\label{cannon}

In this section we consider partial resolutions of the $Y^{p,q}$ singularities
where one blows up an orbifold Fano divisor. These may be described as a K\"ahler
quotient of $\C^5$ by $U(1)^2$ with charges given by
the kernel of the map defined by
\bea
A & = & \left(
\begin{array}{ccccc}
1 & 1 & 1 & 1 & 1 \\
1 & p & p-q-1 & 0 & s \\
0 & p & p-q & 0 & s\\
\end{array}
\right)~,
\label{delzant}
\eea
with FI parameters in an appropriate chamber $C$.
The last column corresponds to the internal point $w_5=(s,s)$ in Figure \ref{troia}.
As we explain, these partial resolutions may be thought of as the total space of the
canonical orbifold line bundle
over a Fano orbifold $M$, which is the exceptional divisor.

Let us begin by defining
\bea
m~=~-p+q+2s~.
\label{defem}
\eea
For $m\geq 0$ we consider as
 basis for the $\C^5 \ // \ U(1)^2$ quotient, obtained from the kernel of (\ref{delzant}),
the charge vectors
\bea
\left(\begin{array}{ccccc}
s &0&s&p-q-2s&-p+q\\
0&s&0&p-s&-p\\
\end{array}\right)
\label{fano1}
\eea
with both FI parameters taken to be positive.
To see what this quotient is, we effectively drop the last column by setting $z_5=0$,
and consider the resulting $U(1)^2$
 quotient of $\C^4$. The first line in (\ref{fano1}) produces
 ${\cal O}_{\mathbb{CP}^1} (p-q-2s) \oplus {\cal O}_{\mathbb{CP}^1} (0)$, with
the fibre of the first factor being $\C/\Z_s\cong \C$.
Indeed, the non-zero charges give rise to a K\"ahler quotient of $\C^3$ by
the $U(1)$ group $(s,s,p-q-2s)$, which gives ${\cal O}_{\mathbb{CP}^1} (p-q-2s)$,
and then the zero charge entry gives the
product of this with $\C$. This may be thought of
as the bundle ${\cal O}_{\mathbb{CP}^1} (p-q-2s) \oplus {\cal O}_{\mathbb{CP}^1} (0)$.
The second row in (\ref{fano1}) then projectivises this bundle via the quotient
$\C^2\setminus \{  0 \} \to \mathbb{WCP}^1_{[p-s,s]}$ on each $\C^2=\C\oplus\C$ fibre.
This space is a  Fano orbifold, which in \cite{np1} was denoted
 \bea\label{fibrearse}
 M & = & K_{\mathbb{CP}^1}^{m/2} \times_{U(1)} \mathbb{WCP}^1_{[p-s,s]}~.
 \eea
In fact, to make contact with \cite{np1} one should set
\bea
s=p-r~,
\eea
a relation which also appears in this reference.
Adding back the $z_5$ coordinate then gives the canonical line bundle
over $M$.
The fact that the charges in (\ref{fano1}) sum to zero guarantees that the first Chern class  of the
total space vanishes and so is Calabi-Yau.
For every $s$ such that $2s>p-q$, which is equivalent to $m>0$, an
explicit Ricci-flat K\"ahler metric, asymptotic to
the cone metric over $Y^{p,q}$, was
constructed in \cite{np1}.
We shall briefly review these metrics in the next section.
\begin{figure}[!ht]
\vspace{5mm}
\begin{center}
\epsfig{file=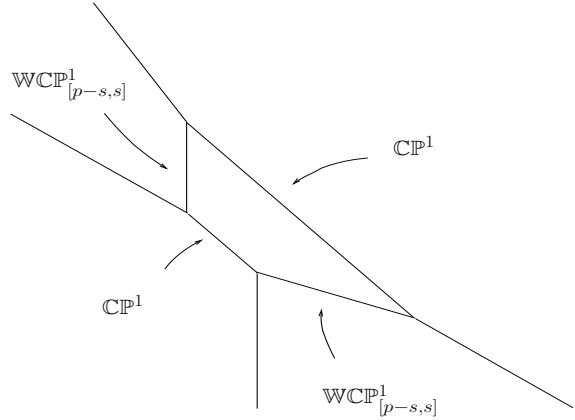,width=7.5cm,height=5.5cm}
\end{center}
\caption{pq-web of a canonical orbifold resolution of a $C(Y^{p,q})$ singularity. The quadrangle
represents the  compact divisor $D_5$, which is the Fano orbifold $M$. The four
non-compact divisors $D_a=\{z_a=0\}$, $a=1,\dots,4$,
are the total spaces of the orbifold line bundles ${\cal O}_{\mathbb{WCP}^1_{[p-s,s]}}(-p)$, ${\cal O}_{\mathbb{CP}^1}(-p-q)$,
 ${\cal O}_{\mathbb{WCP}^1_{[p-s,s]}}(-p)$, ${\cal O}_{\mathbb{CP}^1}(-p+q)$, respectively. Slightly more 
precisely, these are all K\"ahler quotients of $\C^3$ by the $U(1)$ actions with weights $(p-s,s,-p)$, 
$(p-s,p-s,-p-q)$, $(p-s,s,-p)$, $(s,s,-p+q)$, respectively, and with positive moment map level.}
\vspace{5mm}
\label{california}
\end{figure}

The pq-web is drawn in Figure \ref{california}.
The four line segments are  images of copies of $\mathbb{CP}^1$ and $\mathbb{WCP}^1_{[p-s,s]}$.
More precisely, the segments on the left and right hand side of the quadrangle representing
the blown up $M$ are two copies of $\mathbb{CP}^1$. These are orbifold divisors in
$M$, having normal fibres $\C/\Z_{s}$ and $\C/\Z_{p-s}$, and were denoted
$D_2,D_1$ in  \cite{np1}, respectively. The segments at the top and bottom represent
two copies of $\mathbb{WCP}^1_{[p-s,s]}$. The four intersection points are the fixed
points of the $U(1)^3$ action on $X$, and have tangent
cones $\C^3/\Z_{p-s},\C^3/\Z_{s},\C^3/\Z_{p-s},\C^3/\Z_{s}$, respectively (see Figure
\ref{california}). More precisely, in each case the $\Z_n\subset U(1)\subset SU(2)\subset SU(3)$
quotient produces the $\mathcal{N}=2$ $A_{n-1}$ singularity $\C\times\C^2/\Z_n$, where either $n=s$ or $n=p-s$.
These may be deduced from the dual toric diagram -- Figure \ref{troia} -- by applying Delzant's theorem
for each neighbourhood of the four points.

Finally, suppose instead that $m\leq 0$. Consider as
 basis for the $\C^5 \ // \ U(1)^2$ quotient, obtained from the kernel of (\ref{delzant}),
the charge vectors
\bea
\left(\begin{array}{ccccc}
p-s & -p+q+2s&p-s&0&-p-q\\
0&s&0&p-s&-p\\
\end{array}\right)~.
\label{fano2}
\eea
Repeating the same reasoning as above, we see that this GLSM describes the total space of the canonical line bundle over the Fano orbifold
 \bea
 M & = & K_{\mathbb{CP}^1}^{-m/2} \times_{U(1)} \mathbb{WCP}^1_{[s,p-s]}~.
 \eea
Note then that $m<0$ is equivalent to replacing $m$ with $-m$ in
(\ref{fibrearse}) (so that $-m>0$) and interchanging $s$ and $p-s$.
$M$ is an orbifold fibration over $\mathbb{CP}^1$, which may
be thought of as a projectivisation of the bundle
${\cal O}_{\mathbb{CP}^1} (0)\oplus {\cal O}_{\mathbb{CP}^1} (-p+q+2s)$.
This is also the blown up divisor at $z_5=0$. For these cases no explicit
Ricci-flat K\"ahler metric is known.
The pq-web and corresponding discussion of divisors is qualitatively similar to the case
$m\geq 0$.


\section{Supergravity solutions for resolved $Y^{p,q}$ metrics}
\label{themetrics}

In this section we describe Type IIB supergravity solutions that are
dual to various baryonic branches of the $Y^{p,q}$ quiver gauge
theories. These are constructed from Ricci-flat K\"ahler metrics on
partial resolutions of the singular $C(Y^{p,q})$ \cite{np1}, whose
toric description was given in the previous section, together with
the appropriate Green's function. In the next section we will
present the gauge theory duals of these branches.

\subsection{Ricci-flat K\"ahler metrics on $Y^{p,q}$ partial resolutions}

In reference \cite{np1} we constructed explicit asymptotically conical Ricci-flat K\"ahler
metrics on the partial resolutions discussed in the previous section. Indeed,
one of the aims of section \ref{toricsection} was to express the toric geometry
of these partial resolutions in the orbifold fibration language of \cite{np1},
which is how they were naturally constructed in that reference. In this
subsection we briefly present these metrics, in particular determining the explicit
dependence of the metric parameters on the integers
$p,q,s$ of the previous section.

We start by specialising the metrics to the case of interest.
This sets $n=1$ and $V=\mathbb{CP}^1$ with its standard metric, in the notation
of \cite{np1}. The local Ricci-flat K\"ahler metric $g_X$ is then given by
\bea\label{metric}
\pm g_X & = & \frac{y-x}{4X(x)}\diff x^2 + \frac{y-x}{4Y(y)}\diff y^2 + \frac{X(x)}{y-x}\left[\diff\tau +
(1-y)(\diff\psi -\cos\theta\diff \phi)\right]^2 \nonumber\\
&& + \frac{Y(y)}{y-x}\left[\diff\tau + (1-x)(\diff\psi -\cos\theta\diff \phi)\right]^2 +
(x-1)(1-y)g_{\mathbb{CP}^1}~,
\eea
where the metric functions are given by
\bea
X(x) & = & x-1+\frac{2}{3}(x-1)^2+\frac{2\mu}{x-1}\\
Y(y) & = & 1-y-\frac{2}{3}(1-y)^2+\frac{2\nu}{y-1}~.
\label{functions} \eea The $\pm$ sign in (\ref{metric}) depends on
the choice of metric parameters $\mu$ and $\nu$. The
K\"ahler-Einstein metric on $\mathbb{CP}^1$ is the standard one \bea
g_{\mathbb{CP}^1} = \frac{1}{4} (\diff \theta^2 + \sin^2\theta \diff
\phi^2) \eea which obeys $\mathrm{Ric} = 4 g$. As always for a
Ricci-flat geometry, one is free to scale the overall metric by a
positive constant. This may be regarded as the overall size of the
exceptional sets\footnote{We expect that a \emph{general} asymptotically conical
Ricci-flat K\"ahler metric on the canonical partial resolutions
should depend on two independent resolution parameters. However, the
explicit metrics constructed in \cite{np1} depend only on the
overall size of the exceptional Fano orbifold, implying that the
general metric lies outside the ansatz considered in \cite{np1}.}.

Recall from \cite{np1} that the parameter $\nu$ is uniquely fixed in terms of
two integers $p,k$, obeying
\bea
p< k < 2p
\eea
where we have used the fact that the Fano index of $\mathbb{CP}^1$ is $I(\mathbb{CP}^1)=2$.
The integer $k$ of \cite{np1} is related to the $p$ and $q$ of $Y^{p,q}$ via
\bea
k=p+q~.
\eea
Henceforth we adopt the standard $Y^{p,q}$ notation.
The roots $y_i$ of $Y(y)$ may be expressed in terms of
 $p$ and $q$, and are given by quadratic irrationals in $\sqrt{4p^2-3q^2}$.
These obey \cite{paper2}
\bea
y_1+y_2 +y_3 & = & \frac{3}{2}\nn\\
y_1 y_2 + y_1 y_3 + y_2 y_3 & = & 0\nn\\
y_1 y_2 y_3 & = & 3\nu -\frac{1}{2}
\eea
where
\bea
\nu~ = ~\frac{1}{12}+\frac{p^2-3q^2}{24p^3}\sqrt{4p^2-3q^2}~.
\eea
The roots themselves are given by
\bea
y_1 & = & \frac{1}{4p} (2p -3q - \sqrt{4p^2-3q^2})\nn\\
y_2 & = & \frac{1}{4p} (2p +3q - \sqrt{4p^2-3q^2})~.
\eea

In \cite{np1} we showed that, for the Ricci-flat K\"ahler
metrics on the two small partial resolutions of $C(Y^{p,q})$, the
second metric parameter $\mu$ is fixed simply in terms of $\nu$.
In particular, $\mu=-\nu$ for the small partial resolution $I$ of
section \ref{smally1}, whereas $\mu=0$ for the small partial
resolution $II$ of section \ref{smally2}. Note that, in the latter case,
one should take the minus sign in (\ref{metric}).

For the canonical partial resolutions in section \ref{cannon} with $m>0$, the
parameter $\mu$ instead depends on $p$, $q$ and $s$, where $s>(p-q)/2$
determines the exceptional divisor or blow-up vertex. We may easily
determine this dependence as follows. The equation
\bea
x_\pm & = &  \frac{py_1 y_2} {(p-s) y_1 + sy_2}
\eea
determines $x_\pm$ in terms of $p,q,s$.
Using \cite{np1}
\bea
- 2\mu & = & (x_\pm-1)^2 +\frac{2}{3} (x_\pm - 1)^3
\eea
we obtain \bea\label{nicer}
-2\mu & = & \frac{9m^2q^2\left(2p^2-3q^2-p\sqrt{4p^2-3q^2}+mq\right)}
{\left(2p^2-3q^2-p\sqrt{4p^2-3q^2}+3mq\right)^3}~,
\eea
where $m$ was defined in \eqn{defem}.
In particular, note that setting $m=0$ formally gives $\mu=0$, as expected from
the analysis of \cite{np1}. The metrics are defined only for $m>0$.

We now expand the metric (\ref{metric}) near infinity to extract its subleading
behaviour with respect to the
conical metric. This will allow us to make a general prediction for the order parameter which is
turned on in the gauge theory. Following \cite{np1}, we define
\bea
x & = & \mp \frac{2}{3} r^2 ~,
\eea
where the $\mp$ sign is correlated with the $\pm$ sign in (\ref{metric}).
We then have
\bea
\label{submetric}
\pm g_X & = & \diff r^2 +\frac{2}{3} r^2 \left[ \frac{1}{4Y(y)} \diff y^2 +Y(y)\eta^2
+ (1-y)g_{\mathbb{CP}^1} +\frac{2}{3} \left[\diff \tau +(1-y)\eta\right]^2\right] \nn\\
&\pm & \frac{1}{r^2} \Bigg\{
\frac{3}{2}\left(y-\frac{1}{2}\right)\diff r^2+r^2
\Bigg[\frac{y}{4Y(y)} \diff y^2 + y(1-y)g_{\mathbb{CP}^1}+ \nn\\&+&
\!\! Y(y)\left[ 2(\diff \tau +\eta )-y \eta\right] \eta\Bigg] +
 \frac{2}{3}r^2\left(\frac{1}{2}- y\right)[\diff\tau +(1-y)\eta]^2 \Bigg\}+\cdots ~,
\eea
where we have defined the one-form $\eta = \diff \psi -\cos\theta \diff \phi$. The first line is
precisely the Ricci-flat metric on the cone over the Sasaki-Einstein manifold $Y^{p,q}$.
We see that the subleading behaviour is ${\cal O}(r^{-2})$, indicating the presence of a
dimension two operator turned on in the gauge theory \cite{KW2}. Notice that this term is
universal to all metrics, while sub-subleading terms depend for instance on $\mu$.
This behaviour should be reflected by some distinctive property of the field theory.

\subsection{Warped resolved $Y^{p,q}$ metrics}

As discussed in section \ref{spacetravel}, to construct a baryonic
branch solution we must specify a location for the stack of
D3-branes $\mathbf{x}_0\in X$, and subsequently determine the
corresponding Green's function on $(X,g_X)$. In order to preserve
the isometries of the metrics $g_X$, we shall restrict to
$U(1)^3$-invariant Green's functions. The relevant part of the
Laplace operator reads
\bea \sqrt{\det g_X} \Delta & = & \frac{1}{4}\sin\theta\left[
(1-y) \frac{\de}{\de x} \left (q(x) \frac{\de}{\de x}\right)+
(1-x) \frac{\de}{\de y} \left (p(x) \frac{\de}{\de y}\right)\right]\nn\\
& + & \frac{1}{4}(y-x) \frac{\de}{\de \theta} \left (\sin\theta \frac{\de}{\de \theta}\right)~,
\eea
where we have defined
\bea
&& q(x)~ =~ (1-x)X(x) ~=~ -2\mu-(1-x)^2 +\frac{2}{3} (1-x)^3 ~{}\nn\\
&& p(y) ~= ~(1-y)Y(y) ~= ~-2\nu +(1-y)^2 -\frac{2}{3} (1-y)^3 ~.
\eea
One must then solve
\bea
\sqrt{\det g_X}\, \Delta_{\mathbf x} \,G(\mathbf{x},\mathbf{x}_0) & =  & -{\cal C}\, \delta(\mathbf{x}-\mathbf{x}_0)~.
\label{xmen}
\eea
Of course, the differential equation is the same in all cases, while the boundary conditions
depend on the particular class of resolution.
The Green's functions may then be determined by separation of variables,
and written as a formal expansion\footnote{We are suppressing dependence on the location of the D3-branes,
$\mathbf{x}_0$.}
\bea
G (\theta,y,x) & = & \sum_I \Theta_I (\theta) R_I (y) K_I (x)~
\label{frufru}
\eea
where the sum is over some ``quantum numbers'', here collectively denoted $I$, to be determined.
Equation (\ref{xmen}) then reduces to  three decoupled ordinary differential equations
\bea
&&\frac{\diff }{\diff \theta }\left( \sin\theta \frac{\diff \Theta_I }{\diff \theta } \right)
+ a_1\sin\theta \, \Theta_I ~ = ~ 0\label{spheric}\\
&& \frac{\diff }{\diff y}\left(p(y)\frac{\diff R_I}{\diff y }\right)
-(a_2y +a_1-a_2)\, R_I~ = ~ 0\label{ypsy}\\
&& \frac{\diff }{\diff x}\left(q(x)\frac{\diff K_I}{\diff x }\right) +
(a_2x +a_1 -a_2) \, K_I ~ = ~ 0\label{xpsy}
\eea
where $a_1, a_2$ are two integration constants. Here we have dropped the delta functions;
this may be done, provided of course one then imposes the appropriate boundary conditions
on the solutions.

The solutions to the first equation are just ordinary spherical harmonics
 $P_l(\cos\theta)$ (Legendre polynomials), labelled by an integer $l$ through $a_1=l(l+1)$.
Equations  (\ref{ypsy}) and (\ref{xpsy}) (when $\mu\neq 0$),
 are particular cases of the
 Heun\footnote{
When $\mu=0$, which recall corresponds to small partial resolution II, (\ref{xpsy}) reduces to a hypergeometric equation and
the analysis goes through with obvious modifications.} differential
equation, as may be shown by a simple change of variable
\cite{lifted}. In particular, setting \bea z & = & \frac{y- y_1}{y_2
-y_1}~, \eea equation (\ref{ypsy}) may be written in  the canonical
form \bea \frac{\diff^2 R (z)}{\diff z^2} + \left( \frac{\gamma}{z}
+\frac{\delta}{z-1}+\frac{\epsilon}{z-a}\right) \frac{\diff
R(z)}{\diff z}+
 \frac{\alpha\beta z- \lambda}{z(z-1)(z-a)} \, R(z) ~ = ~ 0 ~,
\label{ripper}
\eea
where  the four singular points are $\{0,1,a,\infty\}$ and the five parameters obey the relation
\bea
\alpha+\beta - \gamma -\delta -\epsilon  +1 & = & 0~.
\label{flipflop}
\eea
By comparison, one reads off the following values of the parameters
\bea
&&a = \frac{y_3-y_1}{y_2-y_1}\qquad \quad \gamma=\delta=\epsilon = 1\nn\\
&&\alpha\beta = -\frac{3}{2} a_2 \qquad \quad \lambda =\frac{3(a_2(y_1-1)+a_1)}{2(y_2-y_1)}~,
\eea
with $\alpha+\beta = 2$, and
\bea
a_2 & = & \frac{2}{3} n(n+2)~,\qquad \quad  n \in \mathbb{N} ~,
\eea
following from regularity.

Equation (\ref{xpsy}) may be dealt with in a similar way, with the roots $x_i$ of $q(x)$
replacing the $y_i$ above. We then arrive at
the general expression for the Green's function
\bea
G (\theta,y,x) & = & \sum_{l, n} \Theta_l (\theta) R_{l n} (y) K_{l n} (x)~.
\eea
The sum runs over two positive integers $l,n$, and the dependence on $\mathbf{x}_0$ may be easily
determined in each case, by an analysis near the source, similarly to \cite{KM}.

In fact, as we discussed in section \ref{spacetravel}, existence and uniqueness of the appropriate Green's
functions is guaranteed by general results. In particular, near to
$\mathbf{x}_0$, the Green's function behaves as
$G(\mathbf{x},\mathbf{x}_0)\sim \rho^{-4} (\mathbf{x},\mathbf{x}_0)$, where $\rho$ is
the geodesic
distance from $\mathbf{x}_0$. The warped resolved metric
\bea
g_{10} & = &  G (\mathbf{x},\mathbf{x}_0)^{-1/2} g_{\R^{1,3}} +   G (\mathbf{x},\mathbf{x}_0)^{1/2}   g_X
\eea
then interpolates between AdS$_5\times Y^{p,q}$ at infinity and AdS$_5\times Z$ in the interior,
where here $Z$ is an appropriate orbifold of $S^5$.
In particular, if $\mathbf{x}_0\in \mathbb{CP}^1$ in the small partial
resolution I, we have the ${\cal N}=2$ orbifold $S^5/\Z_p$; if $\mathbf{x}_0$ is at
the north or south
pole of $\mathbb{WCP}^1$ in the small partial resolution II, we
obtain $\mathcal{N}=1$ orbifolds $S^5/\Z_{p+q}$ or $S^5/\Z_{p-q}$, respectively\footnote{Notice that when  $\mathbf{x}_0$ is a generic (non-singular) point on $\mathbb{WCP}^1$, we have the $A_1$ orbifold
$S^5/\Z_2$ when $p+q$ is even, and simply $S^5$ when $p+q$ is odd.}; finally\footnote{In fact, more generally it turns out
that $Z$ may also be a
$Y^{p',q'}$ Sasaki-Einstein manifold, such that $\vol (Y^{p',q'})>\vol (Y^{p,q})$,
although no explicit
metrics $g_X$ are currently known.},
taking $\mathbf{x}_0$
to be a $U(1)^3$-invariant point in a canonical partial resolution, $Z$
is one of the $\mathcal{N}=2$ orbifolds $S^{5}/\Z_{s}$
or $S^{5}/\Z_{p-s}$, with $0<s<p$.


\section{Baryonic branches of $Y^{p,q}$ quiver theories}
\label{VEV1}

We now turn our attention to the $Y^{p,q}$ quiver gauge theories
\cite{toricpaper,Bertolini:2004xf,quiverpaper} and the dual interpretation of the Ricci-flat
K\"ahler partial resolutions of $C(Y^{p,q})$ described in the
previous two sections. Using the results of section \ref{greeny} one
can argue that placing the $N$ D3-branes at a $U(1)^3$-invariant
point $\mathbf{x}_0\in X$, for $X$ one of the toric partial resolutions
discussed, leads to zero VEVs for most of the bifundamental fields
in the $Y^{p,q}$ theory. We thus give generic VEVs to the remaining
fields, in each case, and analyse where the Higgsed theory flows in
the far IR. We find in each case that the theory flows to an
AdS$_5\times (S^5/\Gamma)$ supersymmetric orbifold theory, where the
action of $\Gamma$ on $S^5\subset\C^3$ precisely agrees with the
near-horizon limit of the $N$ D3-branes at the point $\mathbf{x}_0\in X$. We
obtain both $\mathcal{N}=1$ and $\mathcal{N}=2$ orbifolds this way.

We begin in section \ref{ypqquivers} by briefly reviewing the $Y^{p,q}$ quiver
gauge theories. In sections \ref{smallI}, \ref{smallII} and \ref{pacman}
we study the small partial resolutions $I$ and $II$, and the canonical partial resolutions,
respectively, with various choices for the point $\mathbf{x}_0\in X$.

\subsection{$Y^{p,q}$ quiver gauge theories}
\label{ypqquivers}

The $Y^{p,q}$ quiver gauge theories may
be represented by quiver diagrams with $2p$ nodes, each node having gauge
group $U(N)$. For large $N$ these theories were conjectured to flow
to a non-trivial infra-red fixed point that is AdS/CFT dual to Type
IIB string theory on AdS$_5\times Y^{p,q}$, where $Y^{p,q}$ are the
Sasaki-Einstein manifolds constructed 
in references \cite{paper1,paper2}.

\begin{figure}[ht]
  \hfill
  \begin{minipage}[t]{0.48\textwidth}
    \begin{center}
    \epsfxsize=5cm
    \epsfbox{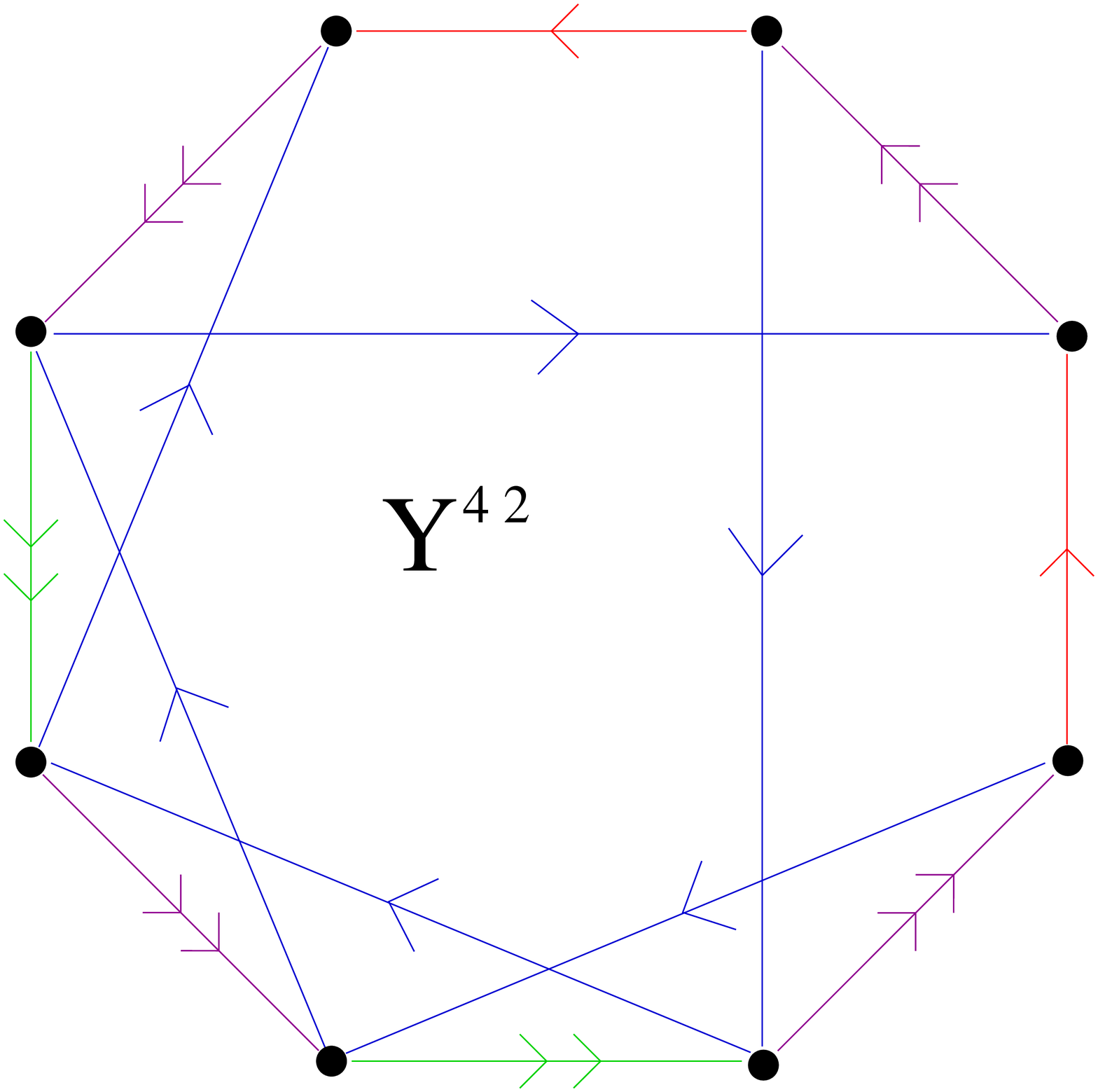}
      \end{center}
  \end{minipage}
  \hfill
  \begin{minipage}[t]{.48\textwidth}
    \begin{center}
     \epsfxsize=5cm
    \epsfbox{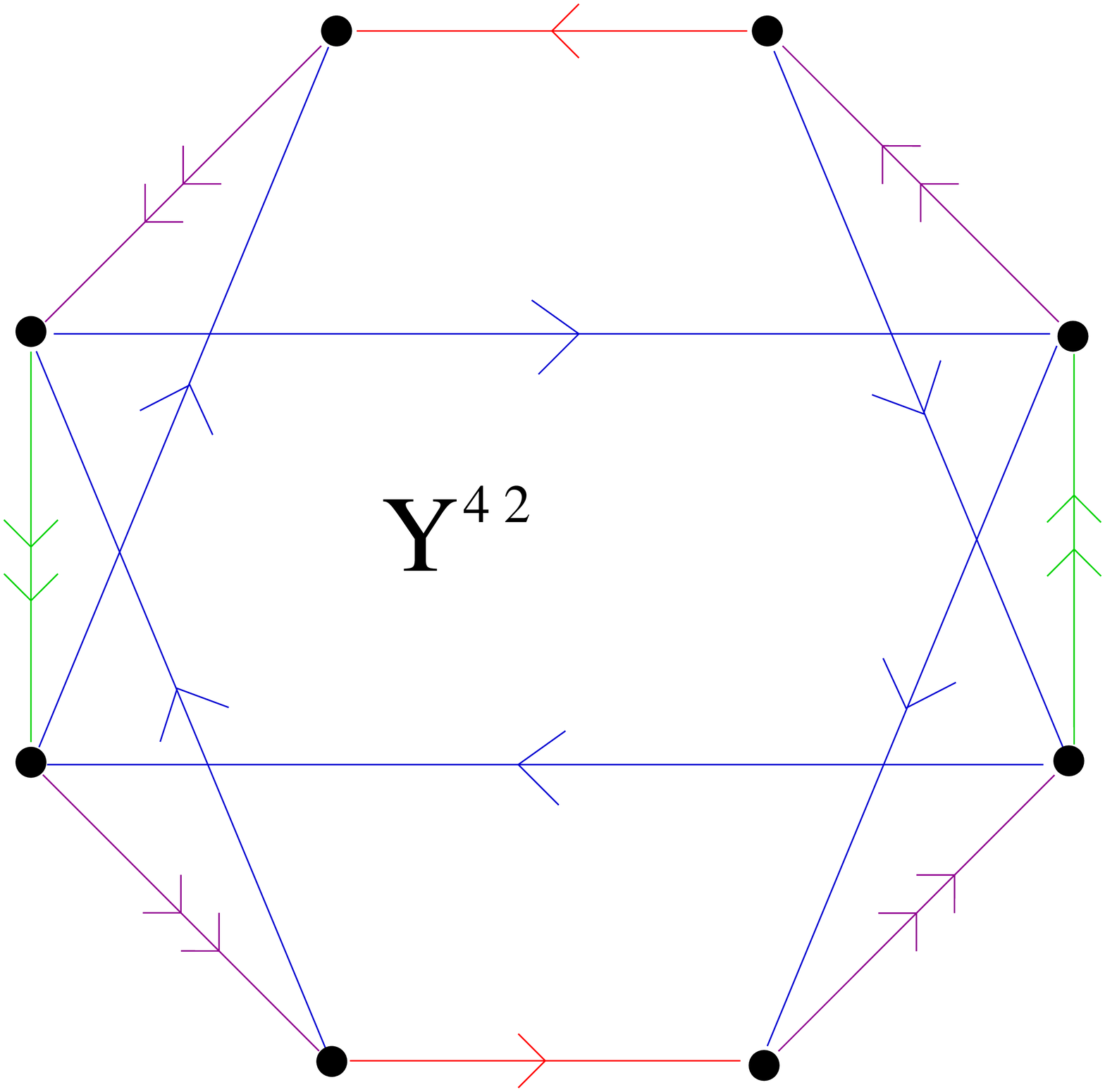}
     \end{center}
  \end{minipage}
\caption{On the left hand side: a $Y^{4,2}$ quiver diagram in the toric phase that we adopt in this
paper. On the right hand side: a $Y^{4,2}$ quiver  in a different
toric phase. The two are related by Seiberg duality.}
  \hfill
  \label{y42}
\end{figure}

The precise field content of a $Y^{p,q}$ theory may be summarised as follows:
\begin{tabbing}
\hspace{1cm} \= $\bullet$ \ \= $p$ $SU(2)$ doublet fields $U^{\alpha}_i$, \ \ \= $i=1,\ldots,p$, \ $\alpha=1,2$ \\
\> $\bullet$ \> $q$ $SU(2)$ doublet fields $V^{\alpha}_i$, \> $i=1,\ldots,q$, \ $\alpha=1,2$ \\
\> $\bullet$ \> $p-q$ $Z_i$ fields, \> $i=q+1,\ldots,p$ \\
\> $\bullet$ \> $p+q$ $Y_i$ fields, \> $i=1,\ldots,p+q$.
\end{tabbing}
In particular, the fields $U^{\alpha}_i$, $V^{\alpha}_i$ are
acted on by an $SU(2)$ flavour symmetry. The representations
under the $2p$ gauge groups may be taken as follows:
\begin{tabbing}
\hspace{1cm} \= $U^{\alpha}_i:$ \ \ \ \= $\overline{\mathbf{N}}_{2i-1}\times \mathbf{N}_{2i}$, \hspace{2.4cm} \= $i=1,\ldots,p$ \\
\> $V^{\alpha}_i:$ \> $\overline{\mathbf{N}}_{2i}\times \mathbf{N}_{2i+1}$, \> $i=1,\ldots,q$
\\
\> $Z_i:$ \> $\overline{\mathbf{N}}_{2i}\times \mathbf{N}_{2i+1}$, \> $i=q+1,\ldots,p$ \\
\> $Y_i:$ \> $\left\{
\begin{array}{l}
  \overline{\mathbf{N}}_{i+2}\times \mathbf{N}_{i}, \qquad \qquad \quad \ \ i=1,\ldots,2q\nn\\[2mm]
 \overline{\mathbf{N}}_{2(i-q)+2}\times \mathbf{N}_{2(i-q)-1}, \quad i=2q+1,\ldots,p+q.
\end{array}\right.$
\end{tabbing}
Here we have introduced, for simplicity, a periodic index $i\in\Z/2p\Z$ for
the nodes of the quiver; thus node $2p+1$ is identified with node 1.
Without loss of generality, we
have chosen a toric phase \cite{phase} for the theory in which all $Z$ fields
appear consecutively in the quiver diagram. For general $p$ and $q$
there exist different
$\mathcal{N}=1$ quiver gauge theories that, via a generalised
form of Seiberg duality, flow to the same infra-red fixed point theory
as the above theories. See Figure \ref{y42} for an example.

The superpotential is constructed from cubic and quartic terms
in the fields, {\it i.e.} closed oriented paths of length three and
four, respectively.
The cubic terms each use one $U,V$ and a $Y$ field of the first kind,
whereas the quartic
terms are constructed using two $U$ fields, one $Z$ and one $Y$ field
of the second kind.
The general superpotential is given by
\bea
W\, =\, \epsilon_{\alpha\beta}\left(\sum_{i=1}^q U_i^\alpha V_{i}^\beta Y_{2i-1}+V_{i}^\alpha U_{i+1}^\beta Y_{2i}\right) - \epsilon_{\alpha\beta}\sum_{i=q+1}^p Z_i U_{i+1}^\alpha Y_{i+q}U_i^\beta~.
\eea
A trace is understood in this formula, and all subsequent such formulae
for $W$.

\begin{figure}[ht]
  \hfill
  \begin{minipage}[t]{0.48\textwidth}
    \begin{center}
    \epsfxsize=5cm
    \epsfbox{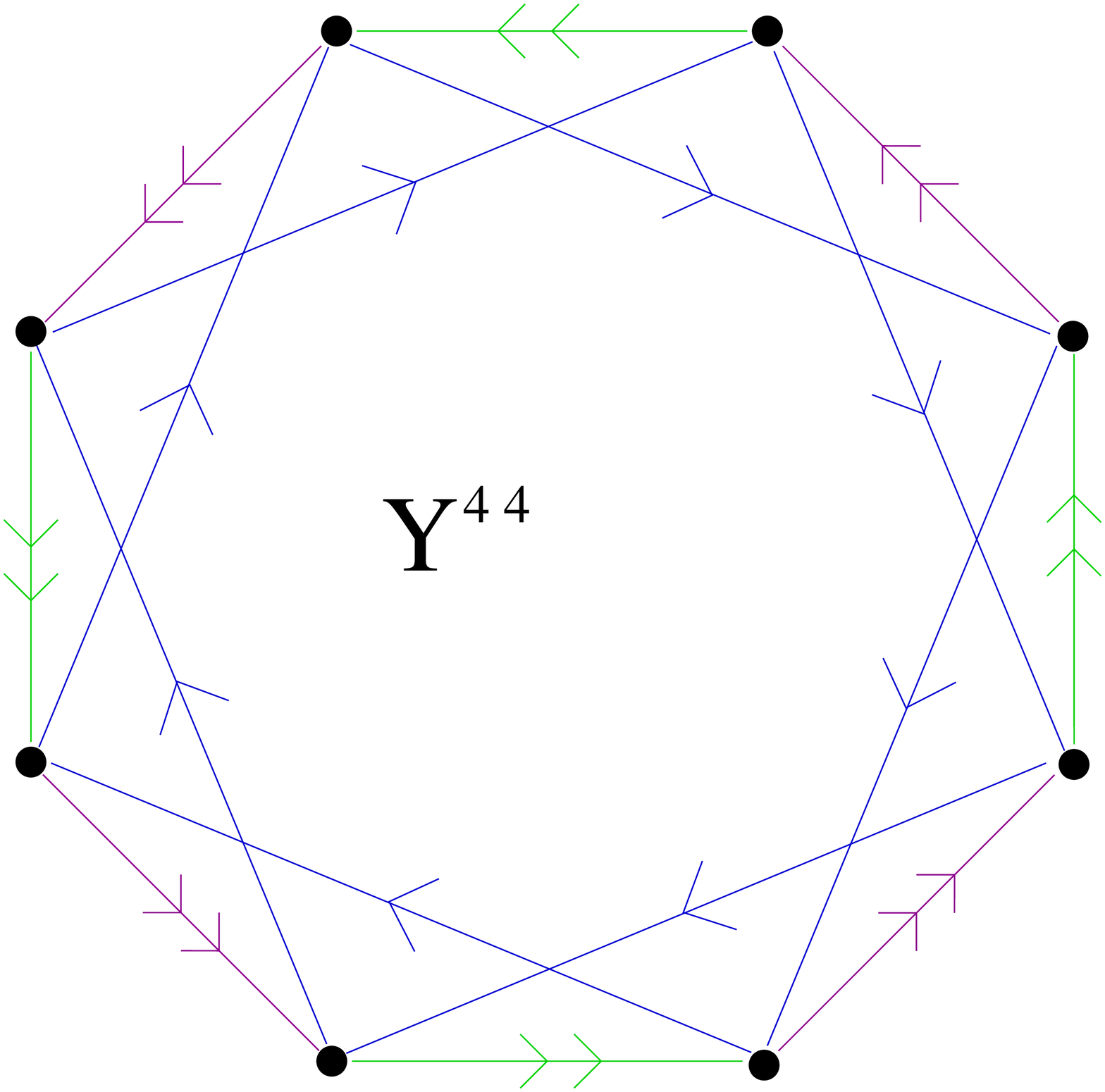}
      \caption{Quiver diagram for $Y^{4,4}$, which is a $\C^3/\Z_4$ orbifold.}
      \label{y44}
    \end{center}
  \end{minipage}
  \hfill
  \begin{minipage}[t]{.48\textwidth}
    \begin{center}
     \epsfxsize=5cm
    \epsfbox{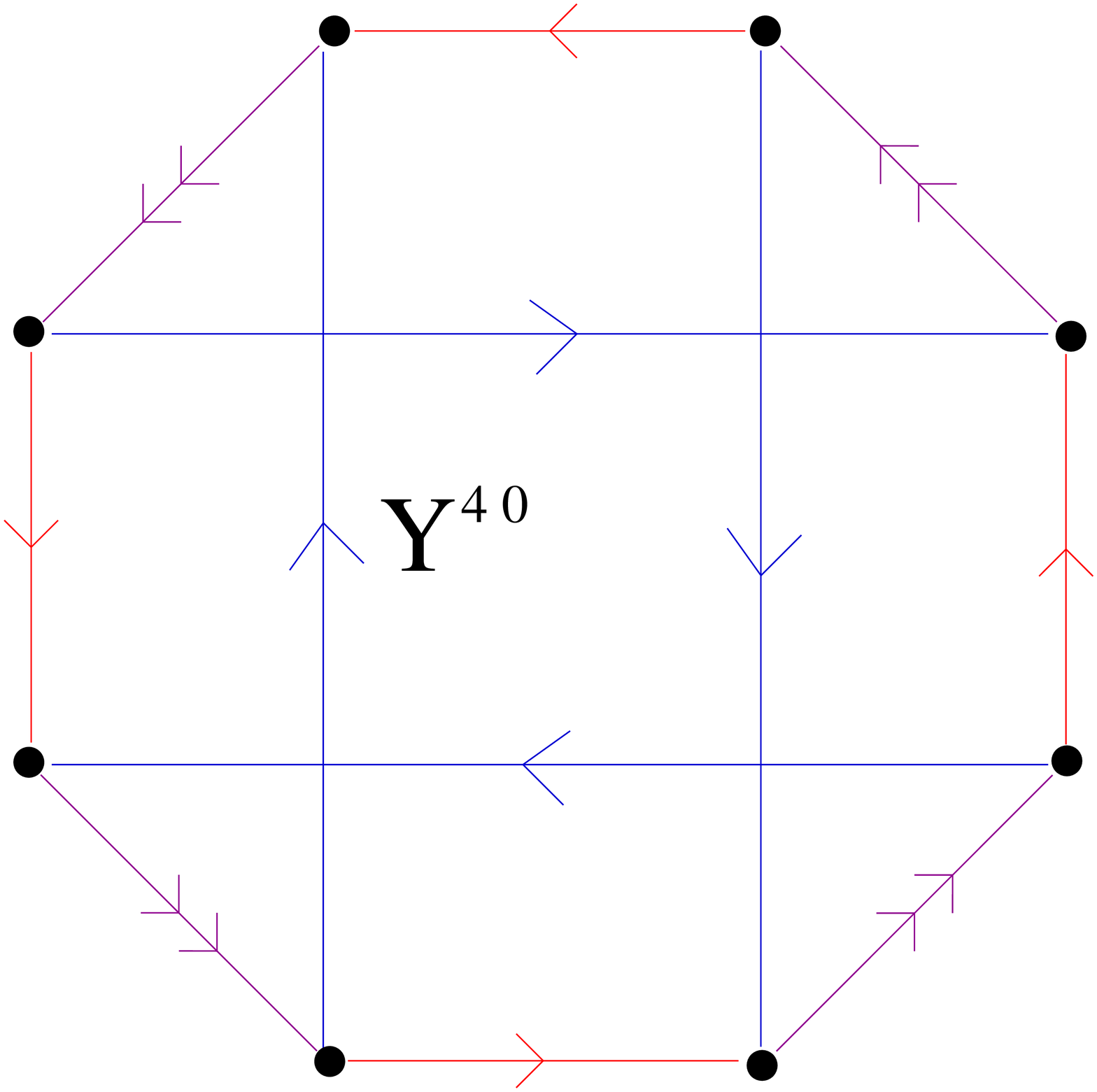}
      \caption{Quiver diagram for $Y^{4,0}$, which is a $\Z_4$ orbifold of the conifold.}
      \label{y40}
    \end{center}
  \end{minipage}
  \hfill
\end{figure}

\begin{figure}[ht]
  \hfill
  \begin{minipage}[t]{0.48\textwidth}
    \begin{center}
    \epsfxsize=5cm
    \epsfbox{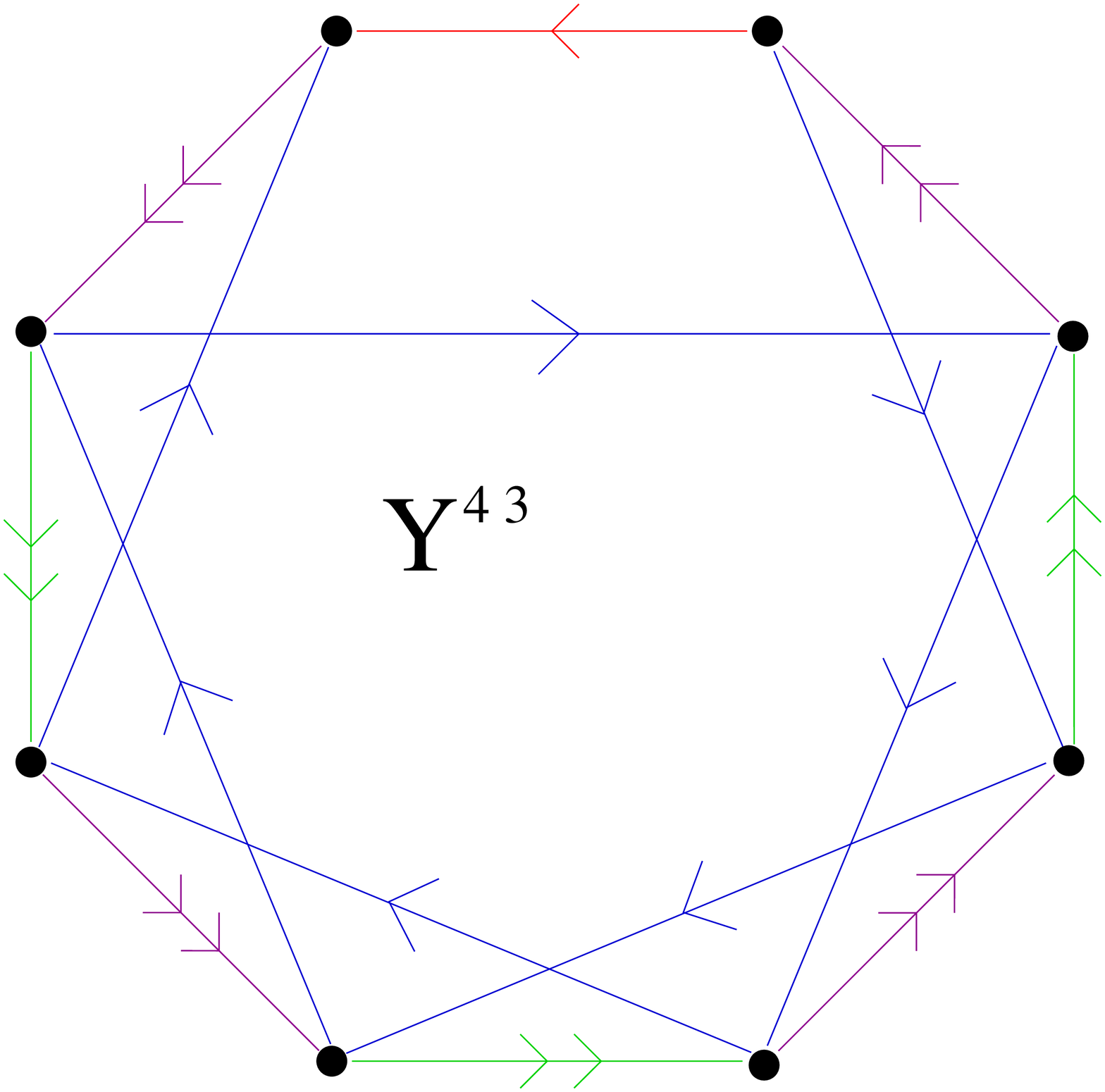}
      \caption{Quiver diagram for $Y^{4,3}$.}
      \label{y43}
    \end{center}
  \end{minipage}
  \hfill
  \begin{minipage}[t]{.48\textwidth}
    \begin{center}
     \epsfxsize=5cm
    \epsfbox{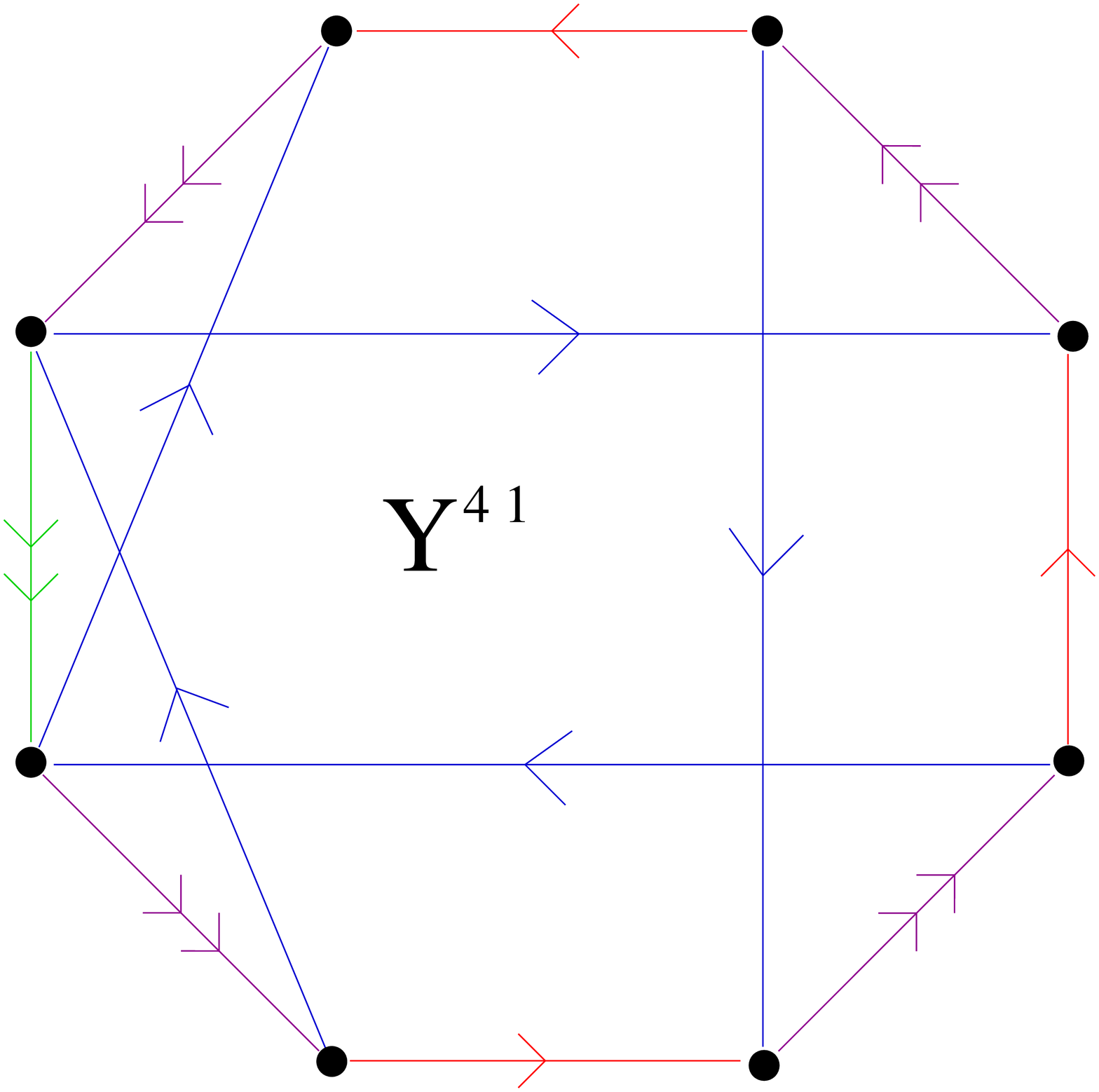}
      \caption{Quiver diagram for $Y^{4,1}$.}
      \label{y41}
    \end{center}
  \end{minipage}
  \hfill
\end{figure}

The $Y^{p,p}$ theories are in fact abelian orbifold quiver gauge theories.
More precisely, they are the orbifold theories obtained by placing
$N$ D3-branes at the origin of $\C^3/\Z_{2p}$ where
the $\Z_{2p}$ group is embedded as $\Z_{2p}\subset U(1)
\subset SU(3)\curvearrowright\C^3$ where the $U(1)$
subgroup of $SU(3)$ is specified by the weight vector $(-2,1,1)$.
The $Y^{p,q}$ theories may then be constructed via an iterative
procedure, described in \cite{quiverpaper}. For illustration, some
quiver diagrams
are shown in Figures  \ref{y44}, \ref{y40}, \ref{y43} and \ref{y41}.
The $U,V,Z$ and $Y$ fields have been colour-coded magenta,
green, red and blue, respectively.

Using the toric description of the $Y^{p,q}$ singularities, to each
toric divisor $D_a$ in the Calabi-Yau cone $C(Y^{p,q})$,
$a=1,\ldots,4$, we may associate baryonic operators ${\cal B}
(\Sigma_a,L_i)$. Here the  $\Sigma_a$, $a=1,\ldots,4$, are the links
of the toric divisors $D_a$, and $L_i$ is a torsion line bundle on
$\Sigma_a$. In the K\"ahler quotient (or equivalently GLSM)
description of $C(Y^{p,q}) = \C^4\ //_0 \ U(1)$ in section \ref{Ypq}
recall that the toric divisors are given by $D_a=\{z_a=0\}$. For
example, we have $\Sigma_1\cong S^3/\Z_p$, so that \bea
|\pi_1(\Sigma_1)|=p~.\eea This leads to $p$ distinct baryonic
particles that may be wrapped on $\Sigma_1$, due to the $p$ distinct
flat line bundles that may be turned on in the worldvolume theory.
In fact the corresponding baryonic operators may be constructed from
determinants of the bifundamental fields $U^1_i$:
\bea \mathcal{B}
(\Sigma_1,L_i) ~ = ~\mathcal{B}(U^1_i) ~=~ \frac{1}{N!}
\epsilon^{\alpha_1\cdots\alpha_N}U_{i,\alpha_1}^{1\, \beta_1} \cdots
U_{i,\alpha_N}^{1\,\beta_N}\epsilon_{\beta_1\cdots\beta_N}~.
\eea
The relation between fields (or rather their corresponding baryons)
and toric divisors for $Y^{p,q}$ is summarised in the table below.
\begin{table}[ht!]
\centerline{
\begin{tabular}{|c|c|c|c|}
\hline
$X_i$ & $\Sigma_a $ & $|\pi_1(\Sigma_a)|$ & $Q_a [U_B(1)]$ \\
\hline
$U^1_i$ & $\Sigma_1$ &  $p$ & $ - p $\\
\hline
$Y_i$ & $\Sigma_2$ &  $p+q $& $p - q$ \\
\hline
$U^2_i$ & $\Sigma_3$ & $ p $& $- p $\\
\hline
$Z_i$ & $\Sigma_4$ &  $p-q$ & $p+q $ \\
\hline
\end{tabular}}
\caption{Bifundamentals of the $Y^{p,q}$ quivers and the corresponding four irreducible toric divisors.}
\label{ypqtable}
\end{table}

The last entry is the baryonic charge, which is precisely the GLSM charge for the
minimal presentation of the singularity \cite{tilings}.
The $V_i^\alpha$ fields, that do not appear in the table,
are slightly more complicated objects from the geometric point of view. These may be associated
to the reducible toric submanifolds $\Sigma_4 \cup \Sigma_1$ and
$\Sigma_3 \cup \Sigma_4$, respectively \cite{tilings}.

Classically, a VEV for a baryonic operator in the UV field theory may be given
by assigning a constant value to a determinant operator, and  this in turn
may be achieved by setting the constituent bifundamental fields to some multiple of the identity matrix.
In other words, giving a VEV to a baryonic operator is, at the classical level, equivalent
to \emph{Higgsing} some of the bifundamental fields.
Therefore, in the following, we will talk about Higgsing
fields or giving VEVs to baryonic operators interchangeably.

The procedure of obtaining new quivers from old ones, via Higgsing
the original theory, is well-studied.
In particular, for toric theories this method allows one to derive, in principle,
a quiver gauge theory that describes the worldvolume theory for $N$ D3-branes
at any partial resolution of the parent toric singularity. 
Although an analysis of the classical moduli space
 of vacua directly from the gauge theory is not available for the $Y^{p,q}$ theories, it is worth noting
 that the following non-chiral protected operator should generically be turned on
 \bea
{\cal U}  =  -p\sum_{i=1}^{p} \sum_{\alpha=1}^2 |U_i^\alpha|^2 +
(p+q)\sum_{i=q+1}^{p}  |Z_i|^2 + (p-q) \sum_{i=1}^{p+q} |Y_i|^2 + q
\sum_{i=1}^{q} \sum_{\alpha=1}^2|V_i^\alpha|^2 ~. \eea This
operator belongs to the conserved baryonic current supermultiplet
of the single non-anomalous baryonic $U(1)_B$ symmetry (recall that
$b_3(Y^{p,q})=1$). This 
has protected conformal dimension $\Delta=2$ and its presence may be
inferred from the subleading expansion of the metrics at infinity
(see section \ref{themetrics}). This is the $Y^{p,q}$ generalisation
of the operator that was originally discussed in \cite{KW2} for the
conifold theory.


\subsection{Small partial resolution I}
\label{smallI}

\begin{figure}[ht]
  \hfill
  \begin{minipage}[t]{0.48\textwidth}
    \begin{center}
      \epsfig{file=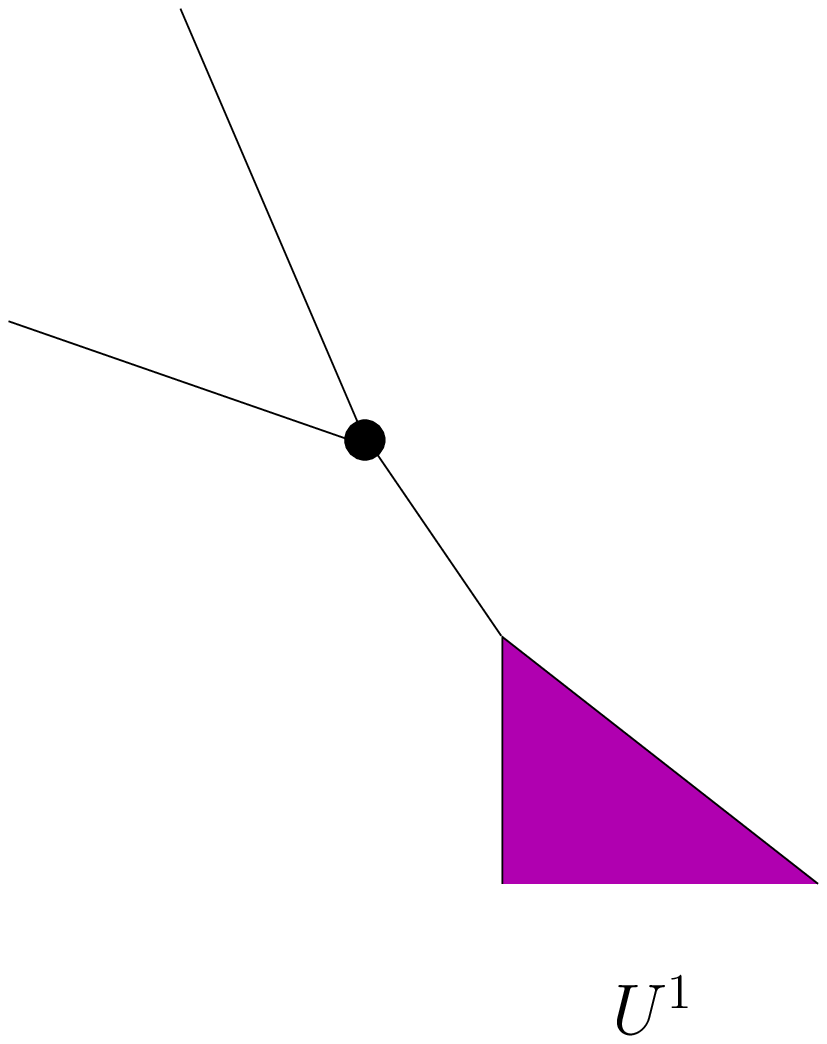, scale=0.5}
     \end{center}
  \end{minipage}
  \hfill
  \begin{minipage}[t]{.48\textwidth}
    \begin{center}
      \epsfig{file=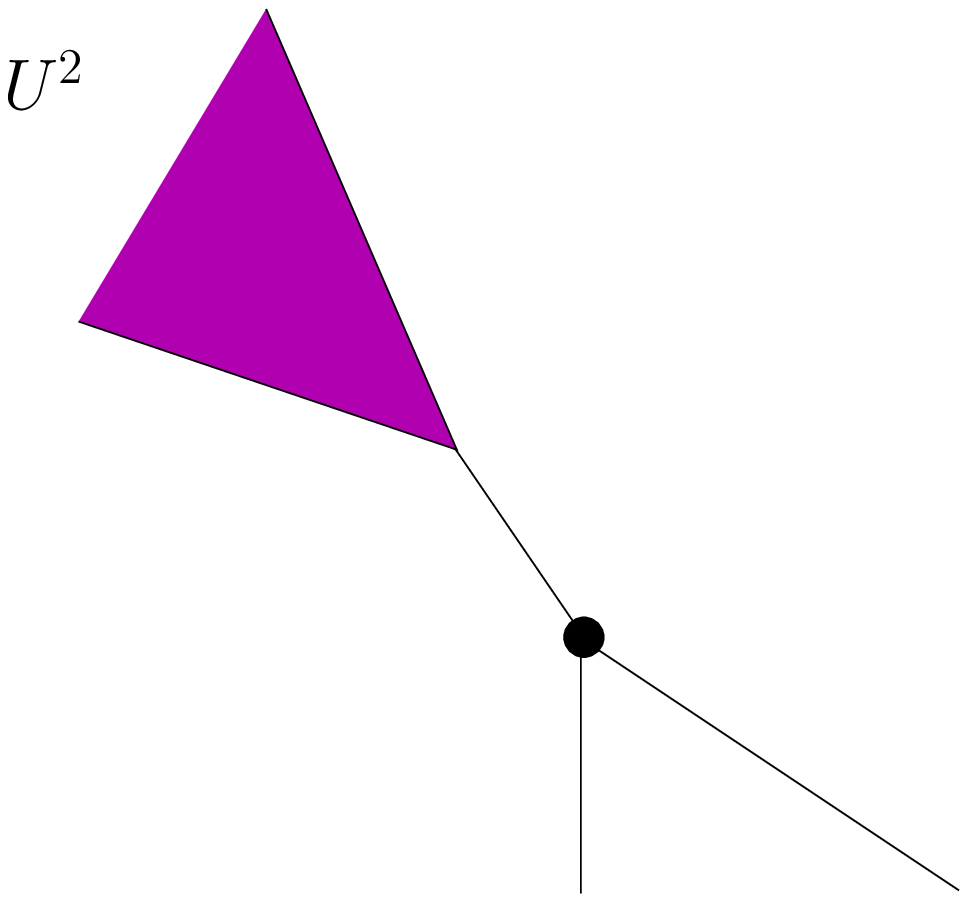, scale=0.5}
     \end{center}
  \end{minipage}
\caption{Placing the D3-branes at the north or south
pole of the exceptional $\mathbb{CP}^1$ in the
first small resolution
   gives VEVs to either the $U^1$ or $U^2$ baryons. These are
related by the action of $SU(2)$.}
  \hfill\label{Uweb}
\end{figure}

\begin{figure}[ht]
  \hfill
  \begin{minipage}[t]{0.48\textwidth}
    \begin{center}
      \epsfig{file=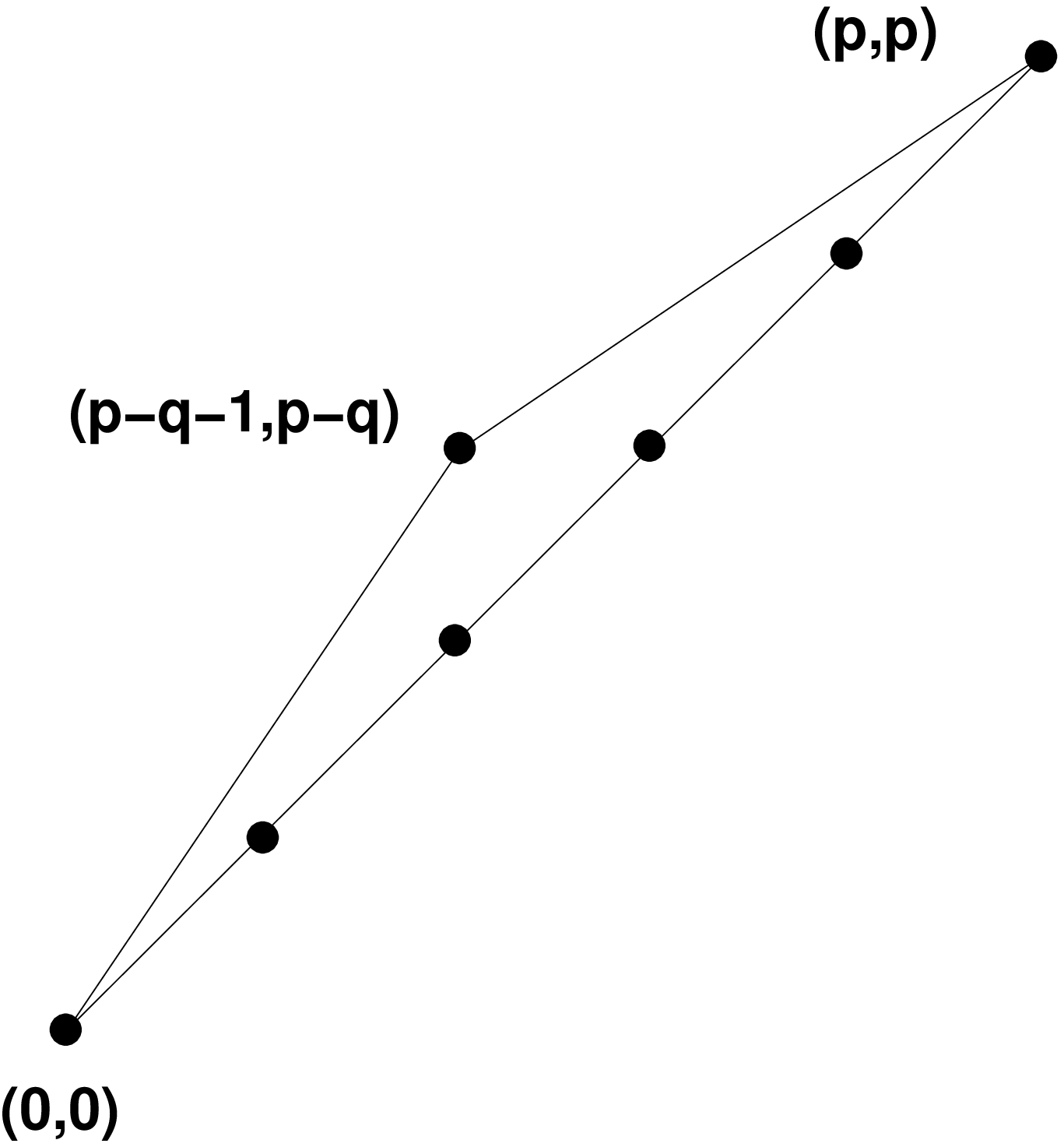, scale=0.3}
    \end{center}
  \end{minipage}
  \hfill
  \begin{minipage}[t]{.48\textwidth}
    \begin{center}
      \epsfig{file=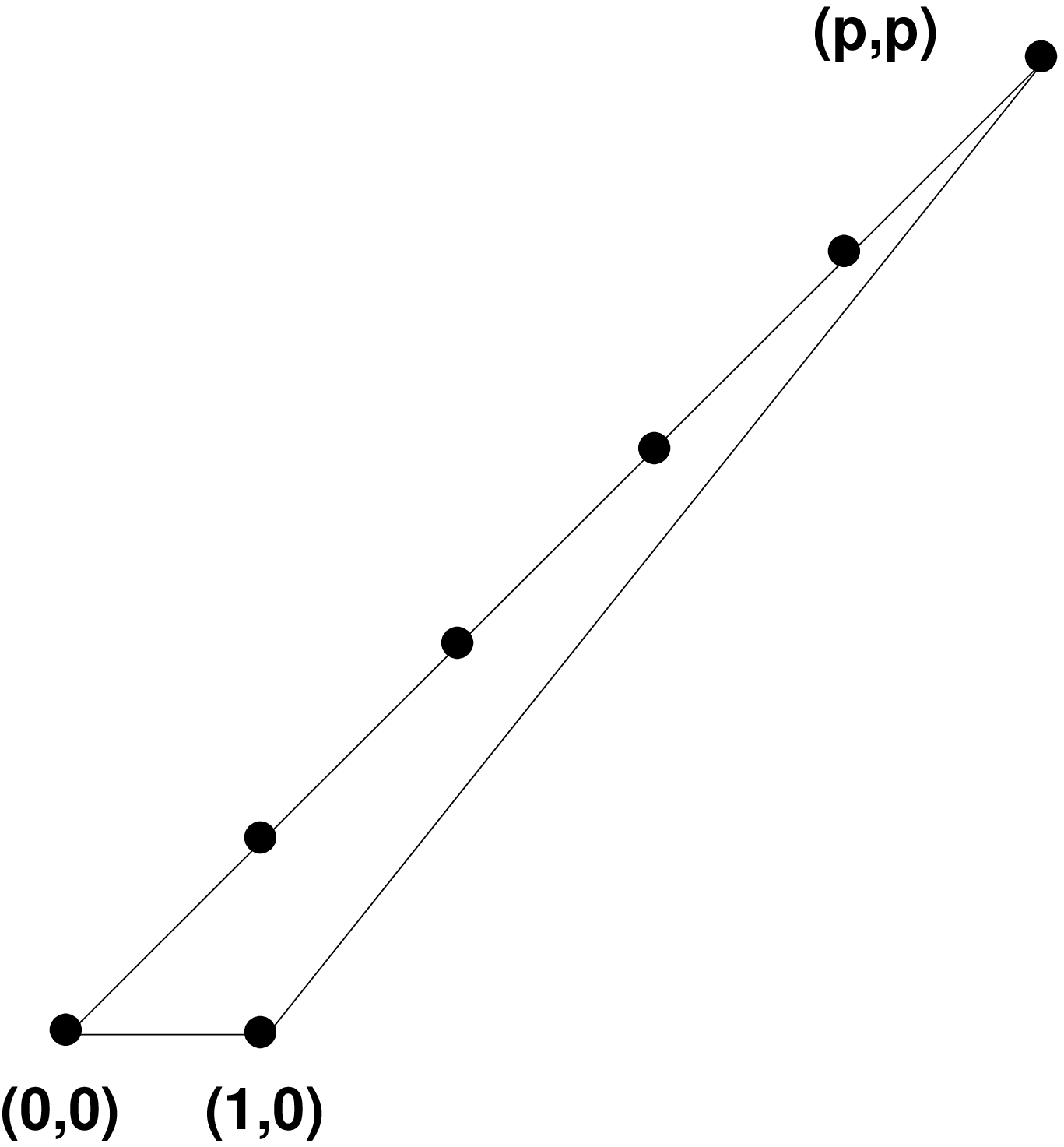, scale=0.3}
      \end{center}
  \end{minipage}
\caption{Toric diagrams for the $A_{p-1}=\C\times\C^2/\Z_p$ orbifold theories
obtained by giving VEVs to the
$U^1$ and $U^2$ baryons, respectively.}
  \hfill\label{Utoric}
\end{figure}

We begin with the small partial resolution I. Consider placing the $N$ D3-branes
at any point $\mathbf{x}_0 \in \mathbb{CP}^1$ on the exceptional $\mathbb{CP}^1$, as shown in
Figure \ref{Uweb}.
All such points are equivalent under the $SU(2)$ isometry of the metric
(\ref{metric}). By placing the $N$ D3-branes at the north (south) pole of $\mathbb{CP}^1=S^2$, the results of section
\ref{greeny} immediately imply that the only toric divisor
that may produce a non-zero condensate is that shaded in Figure \ref{Uweb}. This
corresponds to the fields $U^1_i$ ($U^2_i$), where recall $i=1,\ldots,p$ labels the torsion line bundle.
The theory should flow in the IR to the near horizon geometry of the $N$ D3-branes, which
is determined by the toric diagrams in Figure \ref{Utoric}. Indeed,
according to our general discussion in section \ref{sec3},
this gravity solution should correspond
to an RG flow from the $Y^{p,q}$ theory in the UV to the $\mathcal{N}=2$ $A_{p-1}$ SCFT orbifold theory in the IR,
where the latter arises as the near horizon limit of the branes at the point $\mathbf{x}_0\in \mathbb{CP}^1$.

We now verify these facts directly in the gauge theory.
In particular, we give non-zero VEVs to all $p$ of the $U^1$ fields by setting
\bea\label{U1vevs}
U^1_i = \lambda_i \ \mathrm{I}_{N\times N}\eea
where $\lambda_i\neq 0$ for $i=1,\ldots,p$.
Each chiral matter field is in the bifundamental representation of
$U(N)_i\times U(N)_j$ for the two nodes $i$ and $j$ that
the corresponding arrow connects.
The VEVs (\ref{U1vevs}) then break the gauge symmetry to the diagonal $U(N)$ subgroup.
This breaks the $U(N)^{2p}$ gauge symmetry to $U(N)^p$, where the
nodes of the quiver are effectively contracted pairwise around
the quiver diagram. The VEVs also break the $SU(2)$ flavour symmetry.
The fields $U^2_i$ are adjoints under the
diagonal $U(N)$, and thus become loops at each of the $p$ nodes.
The superpotential becomes
\bea
\widetilde{W} & = & \sum_{i=1}^{q} \lambda_i V_i^2 Y_{2i-1} - U_i^2 V_i^1 Y_{2i-1} +
V_i^1 U_{i+1}^2 Y_{2i} - \lambda_{i+1} V_i^2 Y_{2i}\nn\\
&& + \sum_{i=q+1}^p \lambda_i Z_i U_{i+1}^2 Y_{i+q} - \lambda_{i+1} Z_i Y_{i+q} U_i^2 ~.\eea
Introducing the new fields
\bea
M_i &=& \lambda_i Y_{2i-1}- \lambda_{i+1} Y_{2i}, \qquad i=1,\ldots,q\eea
and substituting for $Y_{2i}$ in terms of $Y_{2i-1}$ one obtains
\bea
\widetilde{W} & = & \sum_{i=1}^{q} V_i^2 M_i -
U_i^2 V_i^1 Y_{2i-1} + \frac{1}{\lambda_{i+1}} V_i^1 U_{i+1}^2 (\lambda_i
Y_{2i-1}-M_i)\nn\\
&& + \sum_{i=q+1}^p \lambda_i Z_i U_{i+1}^2 Y_{i+q} - \lambda_{i+1} Z_i Y_{i+q} U_i^2 ~.\eea
The quadratic terms give masses to the corresponding
fields, which should thus be integrated out in the IR
limit. Integrating out $V^2_i$ sets $M_i=0$ and hence
\bea
\lambda_i Y_{2i-1}= \lambda_{i+1} Y_{2i}\equiv \tilde{Y}_i,\qquad i=1,\ldots,q~.\eea
This reduces the number of $Y$ fields by $q$,
giving $p$ $Y$ and $p$ $\tilde{Y}$ fields in total.
\begin{figure}[ht]
\begin{center}
\epsfig{file=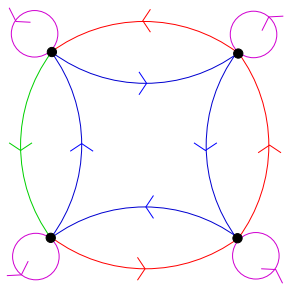, scale=0.32}
\end{center}
\caption{Quiver diagram for the $\C^3/\Z_4$ orbifold theory,
obtained via Higgsing all $U^1$ or all $U^2$ baryons in  the $Y^{4,1}$ theory.
The origin of each field may be traced via its colour.}
\label{Uhiggs}
\end{figure}
Integrating out $M_i$  sets $\lambda_i V_i^2 = \lambda_{i+1} V_i^1U_{i+1}^2$.
In the IR we thus obtain the
effective superpotential
\bea\label{harry}
\widetilde{W}_{\mathrm{eff}} \  = \ \sum_{i=1}^q \frac{1}{\lambda_{i+1}} V_i^1 U_{i+1}^2 \tilde{Y}_i - \frac{1}{\lambda_i} U_i^2 V_i^1 \tilde{Y}_i
+ \sum_{i=q+1}^p \lambda_i Z_i U^2_{i+1} Y_{i+q} - \lambda_{i+1} Z_i Y_{i+q} U_i^2 ~.\eea
This is precisely the matter content, and cubic superpotential,
of the $\mathcal{N}=2$ $A_{p-1}$ orbifold theory. There are $p$ gauge groups
$i=1,\ldots, p$, with the following matter content:
\begin{tabbing}
\hspace{1cm} \= $U^2_i:$ \ \ \ \= $\mathrm{Ad}[U(N)_{i}]$, \hspace{3cm} \= $i=1,\ldots,p$ \\
\> $V^2_i:$ \> $\overline{\mathbf{N}}_{i}\times \mathbf{N}_{i+1}$, \> $i=1,\ldots,q$ \\
\> $Z_i:$ \> $\overline{\mathbf{N}}_{i}\times \mathbf{N}_{i+1}$, \> $i=q+1,\ldots,p$ \\
\> $\tilde{Y}_i:$ \> $\overline{\mathbf{N}}_{i+1}\times \mathbf{N}_{i}$, \> $i=1,\ldots,q$
\\
\> ${Y}_i:$ \> $\overline{\mathbf{N}}_{i-q+1}\times \mathbf{N}_{i-q}$, \> $i=2q+1,\ldots,p+q$.
\end{tabbing}
The final quiver for $Y^{4,1}$ is shown in
Figure \ref{Uhiggs}. Note that the couplings $\lambda_i$ may effectively
all be set equal to one in (\ref{harry}) by the field redefinitions
\bea
U^2_i &=& \lambda_i\tilde{U}^2_i, \qquad i=1,\ldots,p\nn \\
Z_i & =&
\frac{1}{\lambda_i\lambda_{i+1}}\tilde{Z}_i, \quad i=q+1,\ldots,p+q~.\eea

\subsection{Small partial resolution II}
\label{smallII}

In this section we consider the second small partial resolution. There
are various inequivalent points $\mathbf{x}_0\in \mathbb{WCP}^1$ to place the
$N$ D3-branes.

\subsubsection{Higgsing $Z$}
\label{Zall}

\begin{figure}[ht]
  \hfill
  \begin{minipage}[t]{0.48\textwidth}
    \begin{center}
     \epsfig{file=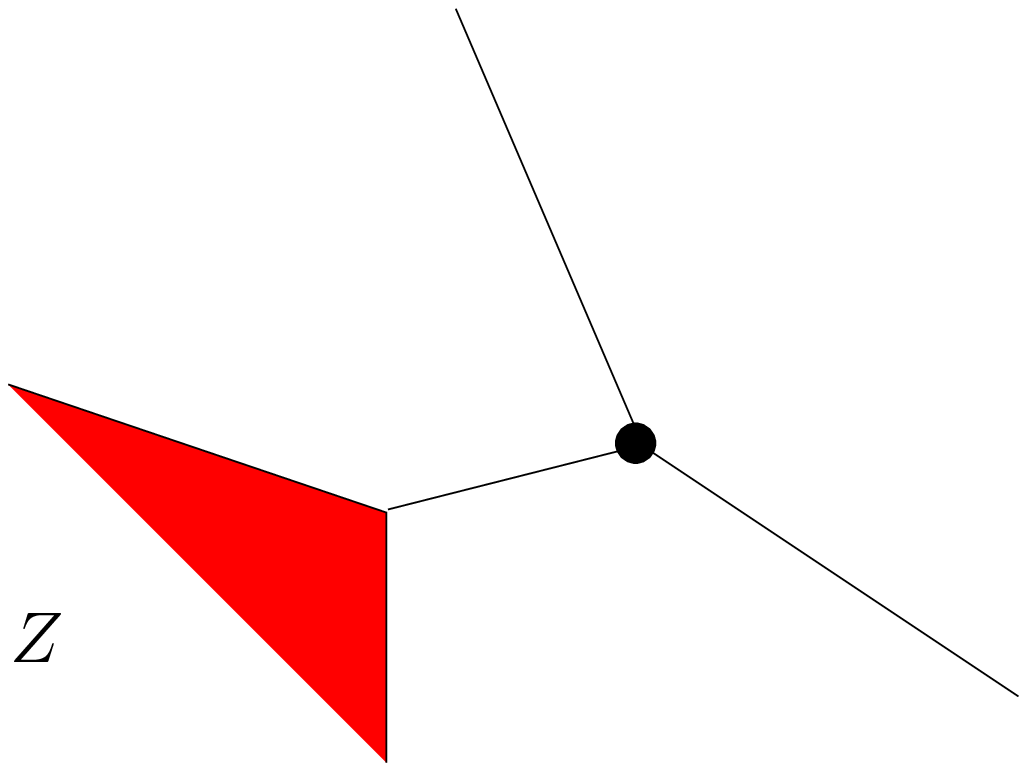, scale=0.5}
    \end{center}
  \end{minipage}
  \hfill
  \begin{minipage}[t]{.48\textwidth}
    \begin{center}
      \epsfig{file=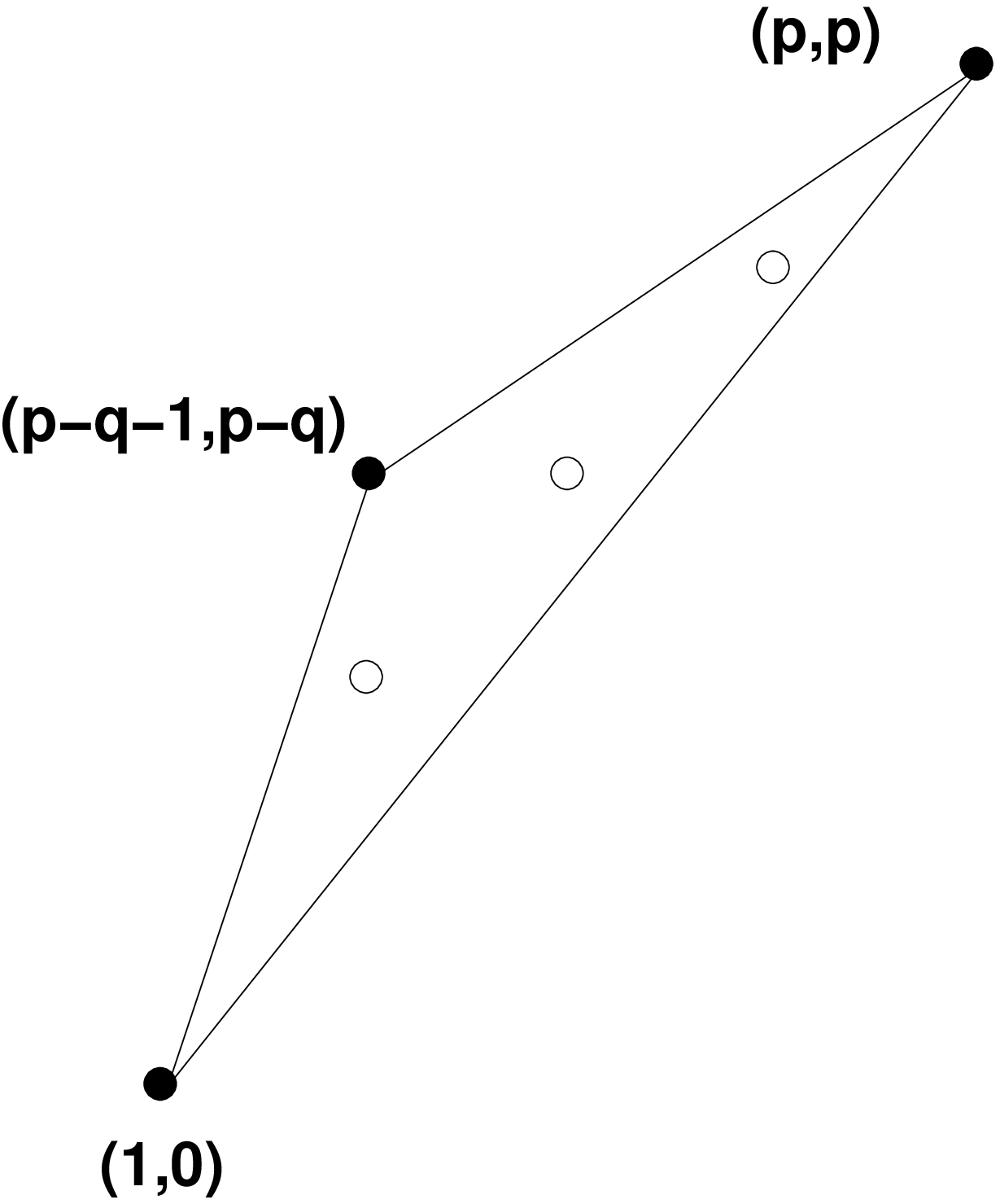, scale=0.3}
    \end{center}
  \end{minipage}
\caption{Placing the D3-branes at the north pole of the exceptional
$\mathbb{WCP}^1$ in the
second small partial resolution gives VEVs to all the $Z$ baryons.}
  \label{Zpic}
  \hfill
\end{figure}

Consider first placing the $N$ D3-branes at the north pole
of the exceptional $\mathbb{WCP}^1$, as shown in Figure \ref{Zpic}.
This point has local geometry (tangent cone) $\C^3/\Z_{p+q}$, where
recall from section \ref{smally2} that the latter is embedded as
$\Z_{p+q}\subset U(1)\subset SU(3)$ where the $U(1)$ subgroup has
weights $(-2,1,1)$. According to our general discussion,
the only fields that may acquire VEVs are the Z fields, and the
theory should flow in the IR to the $\mathcal{N}=1$
orbifold theory corresponding to the abelian quotient singularity
$\C^3/\Z_{p+q}$.

To verify the above directly in the gauge theory, we thus Higgs all $p-q$ of the $Z_i$ fields, $i=q+1,\ldots,p$,
by setting
\bea
Z_i = \lambda_i \ \mathrm{I}_{N\times N}\eea
where $\lambda_i\neq 0$ for $i=q+1,\ldots,p$.
The Higgsing breaks to the diagonal $U(N)$ gauge
groups: this contracts $p-q$ nodes in the quiver pairwise, leaving
a $U(N)^{p+q}$ gauge theory. Of course, since $Z$ is a singlet under $SU(2)$, the VEVs preserve
the $SU(2)$ symmetry. The cubic terms in the superpotential
are unaffected. We obtain the superpotential
\bea
\widetilde{W}_{\mathrm{eff}}=\epsilon_{\alpha\beta}\sum_{i=1}^q U_i^\alpha V_{i}^\beta Y_{2i-1}+V_{i}^\alpha U_{i+1}^\beta Y_{2i} + \epsilon_{\alpha\beta}\sum_{i=q+1}^p \lambda_i U_{i+1}^\alpha {Y}_{i+q}U_i^\beta~.\eea
Note that the couplings $\lambda_i$ may be effectively set to unity
by the field redefinitions $\tilde{Y}_{i+q}=\lambda_i Y_{i+q}$,
$i=q+1,\ldots,p$.
This is indeed precisely the gauge theory for
the $\mathcal{N}=1$ $\C^3/\Z_{p+q}$ orbifold singularity \cite{quiverpaper}.

\begin{figure}[ht]
  \epsfxsize = 4.5cm
  \centerline{\epsfbox{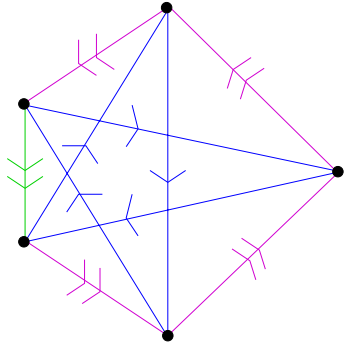}}
  \caption{Quiver diagram for the $\C^3/\Z_5$ orbifold theory, obtained via Higgsing all
  $Z$ baryons in the $Y^{4,1}$ theory. The origin of each field may be traced via its colour.}
  \label{starquiver}
\end{figure}

\subsubsection{Higgsing $Y$}
\label{Yall}

\begin{figure}[ht]
  \hfill
  \begin{minipage}[t]{0.48\textwidth}
    \begin{center}
     \epsfig{file=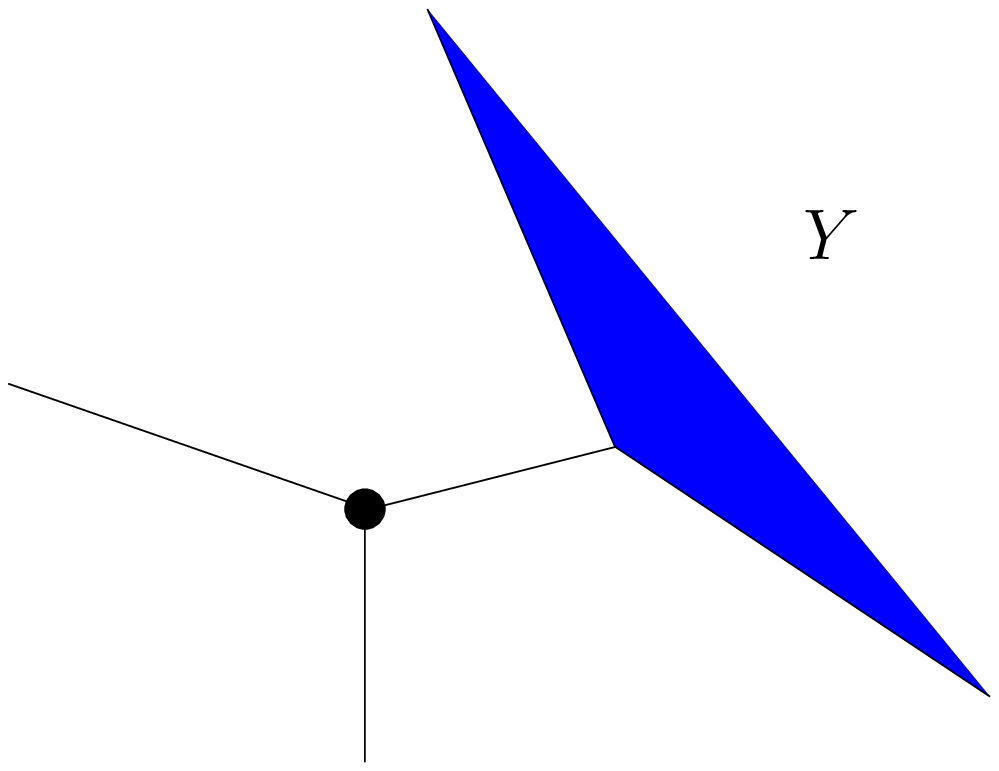, scale=0.5}
    \end{center}
  \end{minipage}
  \hfill
  \begin{minipage}[t]{.48\textwidth}
    \begin{center}
      \epsfig{file=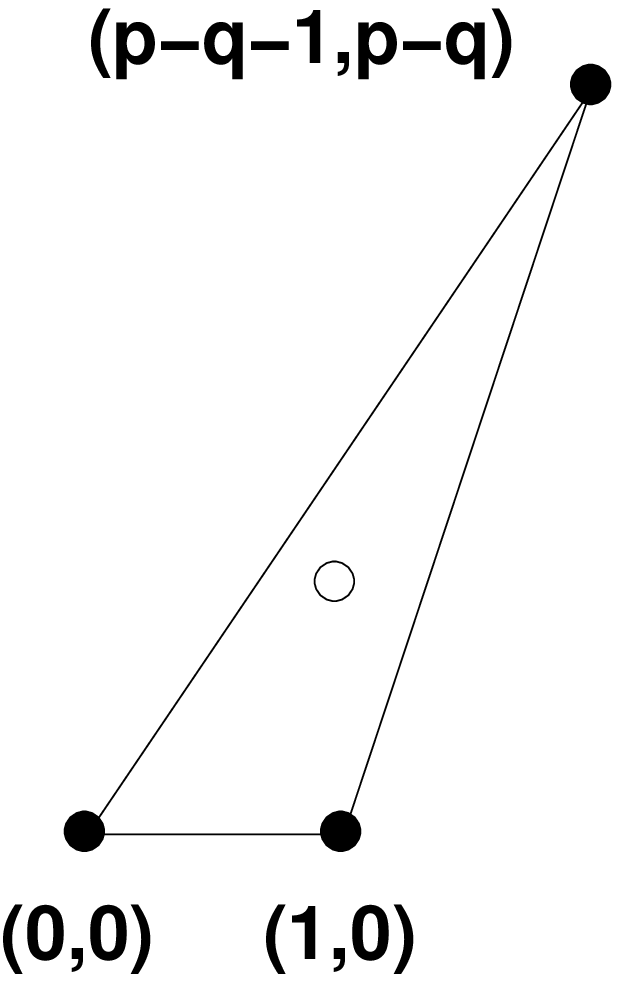, scale=0.3}
    \end{center}
  \end{minipage}
\caption{Placing the D3-branes at the south pole of the exceptional
$\mathbb{WCP}^1$ in the
second small partial resolution gives VEVs to all the $Y$ baryons.}
  \label{Ypic}
  \hfill
\end{figure}

Next consider placing the $N$ D3-branes at the south pole
of the exceptional $\mathbb{WCP}^1$, as shown in Figure \ref{Ypic}.
This point has local geometry (tangent cone)
$\C^3/\Z_{p-q}$, where again the latter is embedded as
$\Z_{p-q}\subset U(1)\subset SU(3)$ where
the $U(1)$ subgroup has weights $(-2,1,1)$. According to our
general discussion,
the only fields that may acquire VEVs are the $Y$ fields, and the
theory should flow in the IR to the $\mathcal{N}=1$
orbifold theory corresponding to the abelian quotient singularity
$\C^3/\Z_{p-q}$.

We thus Higgs all $p+q$ of the $Y_i$ fields, $i=1,\ldots,p+q$,
by setting
\bea
Y_i = \lambda_i \ \mathrm{I}_{N\times N}\eea
where $\lambda_i\neq 0$ for $i=1,\ldots,p+q$.
This Higgsing leaves a $U(N)^{p-q}$ theory, one gauge group for each
$Z$ field.
Again the Higgsing leaves $SU(2)$ unbroken, resulting in the superpotential
\bea\label{hermione}
\widetilde{W} = \epsilon_{\alpha\beta}\sum_{i=1}^q V_{i}^\beta \left(
\lambda_{2i-1}U_i^{\alpha}-\lambda_{2i}U_{i+1}^{\alpha}\right)- \epsilon_{\alpha\beta}\sum_{i=q+1}^p \lambda_{i+q} Z_i U_{i+1}^\alpha U_i^\beta~.
\eea
We make the following field redefinition
\bea
M_i^{\alpha} = \lambda_{2i-1}U_i^{\alpha}-\lambda_{2i}U_{i+1}^{\alpha},
\qquad i=1,\ldots,q\eea
and solve for $U_{i+1}^{\alpha}$, for $i=1,\ldots,q$, in terms of
$U_1^{\alpha}$ and $\{M_i^{\alpha}\}$.
The first sum in (\ref{hermione}) contains only quadratic terms, resulting in masses
for these fields. In particular, integrating out $V_i^{\beta}$ sets
$M_i^{\alpha}=0$ for all $i=1,\ldots,q$, resulting in
\bea
\lambda_{2i-1}U_i^{\alpha} = \lambda_{2i}U_{i+1}^{\alpha}, \qquad i=1,\ldots, q~.\eea
This leaves only $p-q$ independent $U^{\alpha}$ fields,
for each $\alpha=1,2$. Integrating out
$M_i^{\alpha}$ allows one to solve for the $V_i^{\beta}$.
The IR superpotential is then
\bea
\widetilde{W}_{\mathrm{eff}} = \epsilon_{\alpha\beta}\sum_{i=q+1}^p \lambda_{i+q}{Z}_i U_{i+1}^\alpha U_i^\beta~,\eea
where as usual we may redefine $\tilde{Z}_i=\lambda_{i+q} Z_i$, $i=q+1,\ldots,p$ to set the coefficients equal to 1.
This is the correct matter content and superpotential for the $\C^3/\Z_{p-q}$
orbifold theory.
\begin{figure}[ht!]
  \epsfxsize = 1.5cm
  \centerline{\epsfbox{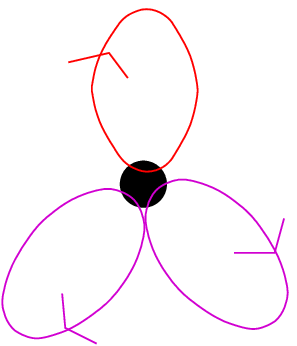}}
  \caption{Quiver diagram for the $\C^3$ theory (${\cal N}=4$ SYM), obtained via Higgsing all
  $Y$ baryons in the $Y^{4,3}$ theory. The origin of each field may be traced via its colour.}
  \label{flowerquiver}
\end{figure}

\subsubsection{Higgsing $Z$ and $Y$}

\begin{figure}[ht!]
 \epsfxsize = 5cm
\centerline{\epsfbox{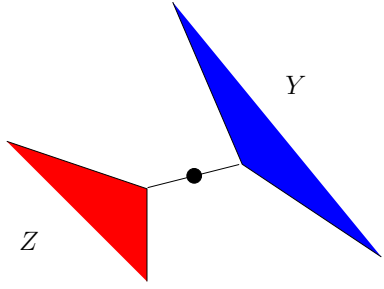}}
\caption{Placing the D3-branes at a generic point on the exceptional $\mathbb{WCP}^1$ of the
second small resolution gives VEVs to all $Z$ and $Y$ baryons.}\label{ZYfigfukky}
\end{figure}

Finally, consider placing the D3-branes at a generic (non-singular) point on the exceptional
$\mathbb{WCP}^1$, as shown in Figure \ref{ZYfigfukky}. The near horizon
limit of the branes depends on the parity of $p+q$: for
$p+q$ even one obtains $\C\times \C^2/\Z_2$, whereas for $p+q$ odd
one obtains $\C^3$. Thus this gravity solution describes
an RG flow from the $Y^{p,q}$ theory in the UV to either the
$\mathcal{N}=2$ $A_1$ orbifold theory, for $p+q$ even, or
$\mathcal{N}=4$ SYM, for $p+q$ odd. Note that only in the
former case is there an explicit Ricci-flat K\"ahler metric in section \ref{themetrics}.

The picture in Figure \ref{ZYfigfukky} suggests that we Higgs all the
$Z$ and $Y$ baryons simultaneously. We thus give the following non-zero VEVs:
\bea
Y_i & = &  \lambda_i \ \mathrm{I}_{N\times N} + \tilde Y_i~,  \qquad  i=1,\ldots,p+q \nn\\
Z_i & = & \mu_i \ \mathrm{I}_{N\times N} + \tilde Z_i~,  \qquad  i=q+1,\ldots,p~.
\label{vevvy}
\eea
Notice that we have included explicitly the fluctuation fields around the vacuum expectation values.

Recalling that the quiver is in a toric phase where all loops corresponding to
cubic and quartic superpotential terms appear consecutively on going around the quiver, one can verify that
starting from any node of the quiver, and grouping it with gauge groups (nodes) connected to
the first one by a Higgsed field ($Z$ or $Y$), there are two possibilities:
(1) if  $p+q$ is even, the nodes are divided into two disjoint sets of $p$
gauge groups each, and therefore the unbroken gauge groups are the
two diagonal subgroups respectively, which we denote $U(N)_1\times U(N)_2$,
(2) if $p+q$ is odd, chasing around the quiver the nodes that are connected by fields that have  a
non-zero VEV, we see that all nodes are covered. Thus the
unbroken gauge group is simply the diagonal $U(N)_\mathrm{diag}$.

Most of the calculation of the effective IR superpotential may be carried out
 for the two cases simultaneously,
and we will indicate at which point the two calculations differ. Inserting (\ref{vevvy})
into the superpotential one obtains
\bea
 \widetilde W &=&
\epsilon_{\alpha\beta}  \sum_{i=1}^q V_i^\beta (\lambda_{2i-1}U_i^\alpha -
\lambda_{2i} U_{i+1}^\alpha)
- \epsilon_{\alpha\beta}\sum_{i=q+1}^p \mu_i \lambda_{i+q}
U_{i+1}^\alpha  U_i^\beta\nn\\
&+ & \eab\sum_{i=1}^q \left( U_i^\alpha V_i^\beta \tilde Y_{2i-1} + V_i^\alpha
U_{i+1}^\beta\tilde Y_{2i} \right)\nn \\
&- &  \epsilon_{\alpha\beta}\sum_{i=q+1}^p \left(\mu_i  U^\alpha_{i+1} U_i^\beta \tilde Z_i
 +\lambda_{i+q} U^\beta_i  U_{i+1}^\alpha \tilde Y_{i+q}\right)~,
  \eea
where we have omitted the quartic terms that will turn out to be irrelevant in the IR.
The first line is quadratic in the $2(p+q)$  fields $U$ and $V$; however, not all these fields get masses.
To see how many of them remain massless one must diagonalise the $2(p+q)\times 2(p+q)$
quadratic form in the $U$ and $V$ fields.
It turns out that four linear combinations are massless if $p+q$ is even, whereas only
two combinations are massless if $p+q$ is odd. We may set
$M_i^{\alpha} = \lambda_{2i-1}U_i^{\alpha}-\lambda_{2i}U_{i+1}^{\alpha}$ for
$i=1,\ldots,q$,  and go to the  basis consisting of $M_i^\alpha, V_i^\alpha$ for $i=1,\dots,q$,
and $U_{q+2}^\alpha,\dots, U_p^\alpha,U_1^\alpha$, where we have solved for $U_{q+1}^\alpha$ in terms of the other
fields as\footnote{The constants $c,a_i$ may be determined iteratively in terms of the $\lambda_i$. It
is straightforward, if cumbersome, to write them down.}
\bea
U_{q+1}^\alpha & = & c U_1^\alpha - \sum_{i=1}^q a_i M_i^\alpha~.
\eea
Integrating out $V_i^\alpha$ and $M_i^\alpha$ then implies
\bea
\begin{array}{rcl}
\lambda_{2i-1} U_{i}^\alpha &=&  \lambda_{2i} U_{i+1}^\alpha \\[1.5mm]
V_i^\alpha/a_i& = & \lambda_{2q+1} \mu_{q+1} U_{q+2}^\alpha
\end{array}\qquad i=1,\dots,q~,
\label{cheese}
\eea
respectively. Integrating out the remaining $U^\alpha_i$ fields yields
\bea
\mu_{i-1}\lambda_{q+i-1} U^\alpha_{i-1} & = & \mu_{i}\lambda_{i+q} U^\alpha_{i+1} \qquad i =
q+2,\dots,p\label{puppy}\\
 \mu_{p}\lambda_{p+q} U^\alpha_{p} & = & c \mu_{q+1}\lambda_{2q+1} U^\alpha_{q+2} \qquad (i=1)~.
\label{fishy}
\eea

\begin{figure}[ht]
  \epsfxsize = 4.5cm
  \centerline{\epsfbox{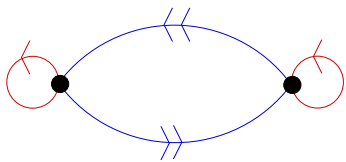}}
 \caption{Quiver diagram for the $A_1$ theory, obtained via Higgsing all $Z$ and $Y$ baryons
 in a $Y^{p,q}$ theory with $p+q$ even. 
 Bifundamentals arise  as a mixture of $U$ and $V$ fields, while the adjoints arise
 as a combination of $Z$ and $Y$ fields.}
\label{a1fig}
\end{figure}

If $p+q$ is even we obtain the following identifications:
\bea
A^\alpha = \frac{U_1^\alpha}{c_1} = \dots = \frac{U_q^\alpha}{c_q} = \frac{U_{q+1}^\alpha}{c_{q+1}}
 = \dots = \frac{U_{p-1}^\alpha} {c_{p-1}} &\in & \mathbf{\overline N}_{1} \times
\mathbf{N}_2~{\;\;}\nn\\
B^\alpha = \frac{V_1^\alpha}{a_1} = \dots = \frac{V_q^\alpha}{a_q} =
\frac{U_{q+2}^\alpha}{c_{q+2}} = \dots =  \frac{U^\alpha_p}{c_p} & \in & \mathbf{N}_{1}
\times \mathbf{\overline N}_2~,
\label{Xmen}\eea
where $c_i$ are constants that may be determined iteratively using the relations
\eqn{cheese} - \eqn{fishy}.
Inserting these into $\widetilde W$, we get the final expression for the effective superpotential
\bea
\widetilde W_\mathrm{eff} & = &   \tilde H_1 \left(A^1 B^2 - A^2 B^1
\right) + \tilde H_2 \left(B^1 A^2 - B^2 A^1\right)
\label{kant}
\eea
where we have defined the two adjoint fields
\bea
\tilde H_1 & = & \sum_{i=1}^{q} a_i c_i \tilde Y_{2i-1} + \sum_{i=1}^{(p-q)/2} c_{q+2i-1}
 c_{q+2i} \left(\lambda_{2q+2i-1} \tilde Y_{2q+2i-1} - \mu_{q+2i-1}\tilde Z_{q+2i-1}\right)\nn\\
\tilde H_2 & = & \sum_{i=1}^{q} a_i c_i \tilde Y_{2i} + \sum_{i=1}^{(p-q)/2} c_{q+2i}
 c_{q+2i+1} \left(\lambda_{2q+2i} \tilde Y_{2q+2i} - \mu_{q+2i}\tilde Z_{q+2i}\right)~.
\eea
This is indeed the correct superpotential for the ${\cal N}=2$  $A_1$ theory,
 as pictured in Figure \ref{a1fig}.

If  $p+q$ is odd, the last entries in the relations \eqn{Xmen} are exchanged, hence $U^\alpha_{p}\sim
U^\alpha_{q+1}$ and $U^\alpha_{p-1}\sim
U^\alpha_{q+2}$, so that all fields get identified on using \eqn{fishy}.
This case may be obtained formally  from the result above, on setting
$X^\alpha = A^\alpha = B^\alpha$
and inserting this into (\ref{kant}). Of course, one has to remember that the gauge group is
broken further
to the diagonal $U(N)_\mathrm{diag}$.
The final expression for the effective superpotential is simply
\bea \widetilde W_\mathrm{eff} & = & \tilde H(X^1
X^2 - X^2 X^1 )~, \eea
where $\tilde H = \tilde H_1 + \tilde H_2$. This is the ${\cal N}=4$ SYM theory,
as expected\footnote{We remark that there are many more Higgsing patterns
that one may consider, resulting in different partial resolutions. Here we have considered a set of examples
motivated by the existence of the corresponding explicit Ricci-flat K\"ahler metrics \cite{np1},
in which the theory
always flows to an orbifold theory in the IR. However, there also exist baryonic
branches where the theory flows between two non-orbifold SCFTs. Rather simple examples may be given
for the $Y^{p,q}$ theories. In particular, giving VEVs to (any) set of  $2s\leq p-q$ $Z$ baryonic operators,
the theory flows to a $Y^{p-s,q+s}$ quiver in the IR. Furthermore, giving VEVs to $2r \leq 2q$ pairs of
$Y$ baryonic operators, the theory flows to a $Y^{p-r,q-r}$ quiver in the IR. In both cases, it may be verified that
the IR value of the $a$ central charge is smaller than that in the IR.}.

\subsection{Canonical partial resolutions}
\label{pacman}

Finally, we consider the canonical partial resolutions of
section \ref{cannon}. These correspond to blowing up
a toric Fano orbifold $M$. The partial resolution is the total space
of the canonical orbifold line bundle over this Fano orbifold. There
are $p-1$ such partial resolutions, labelled
naturally by an integer $s$, with $0<s<p$, that labels
the blow-up vertex in the toric diagram -- see Figure \ref{troia}.
In this section we consider placing the $N$ D3-branes
at the toric fixed points of the exceptional divisor.
As one can see from Figure \ref{california}, there
are four such points. However, two points that lie on the
same $\mathbb{CP}^1\subset M$ divisor in $M$ are related by the isometric
action of $SU(2)$. Thus there are really only two cases to consider.
We consider these in the next two subsections.

\subsubsection{Higgsing $U^1$, $V^1$ and $Z$}

\begin{figure}[ht!]
\begin{center}
\epsfig{file=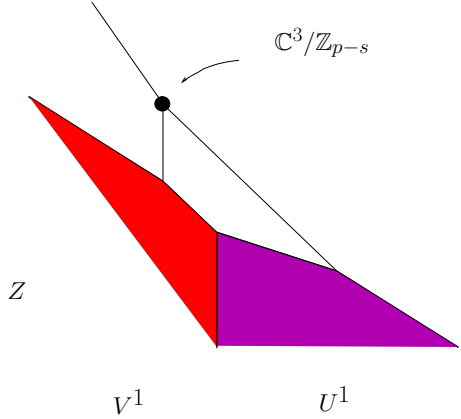,width=6cm,height=5.5cm}
\end{center}
\caption{Placing the D3-branes at the $U(1)^3$-invariant point on the exceptional
divisor, as shown, gives VEVs to a set of $U^1$, $V^1$ and $Z$ baryons.}
\label{UVZvev}
\end{figure}

Placing the D3-branes at the corner of the exceptional divisor
$M$, as in Figure \ref{UVZvev},
implies that no $Y$ or $U^2$ fields
get non-zero VEVs. In this section we show that a certain two-parameter
family of VEVs all flow to the same IR theory, namely
the $\mathcal{N}=2$ $A_{p-s-1}$ orbifold theory. This
is precisely as expected from the gravity dual, since this
is indeed the near-horizon geometry of the stack of D3-branes.

We give the following VEVs:
\bea
U^1_i & = &\lambda_i \ \mathrm{I}_{N\times N}, \qquad i=1,\ldots,p\nn\\
V^1_i & = &\mu_i\ \mathrm{I}_{N\times N}, \qquad i=1,\ldots t\nn\\
Z_i & = & \mu_i\ \mathrm{I}_{N\times N}, \qquad i=q+1,\ldots,q+s-t~.\eea
This is not the most general set of VEVs we could turn on, but
an analysis of the most general case would be too cumbersome;
the above choice for the VEVs is nonetheless still rather general. Here $0\leq t\leq s$,
with $0<s<p$. We also assume, again for simplicity, that $t<q$
and $s-t<p-q$; the non-strict forms of these inequalities \emph{must} hold,
since {\it e.g.} there are only $q$ $V^1$ fields to give VEVs to.
The strict inequalities slightly simplify some of the following
analysis.

The superpotential becomes
\bea
\widetilde{W} &= &\sum_{i=1}^t \lambda_i V_i^2 Y_{2i-1} -
\mu_i U_i^2 Y_{2i-1} + \mu_i U_{i+1}^2 Y_{2i} - \lambda_{i+1}V_i^2 Y_{2i}\nn \\
&+ &\sum_{i=t+1}^q \lambda_i V_i^2 Y_{2i-1} - U_i^2 V_i^1 Y_{2i-1}
+ V_i^1U_{i+1}^2 Y_{2i} - \lambda_{i+1}V_i^2 Y_{2i}\nn\\
&+&\sum_{i=q+1}^{q+s-t} \lambda_i \mu_i U_{i+1}^2 Y_{i+q} -
\lambda_{i+1}\mu_i U_i^2 Y_{i+q} \nn\\
&+& \sum_{i=q+1+s-t}^p \lambda_i
Z_iU_{i+1}^2Y_{i+q} - \lambda_{i+1}Z_i Y_{i+q}U_i^2~.\eea
We introduce the following new fields
\bea\label{rels}
M_i &= &\lambda_iY_{2i-1} - \lambda_{i+1}Y_{2i}, \qquad i=1,\ldots,q\nn\\
N_i & = & \mu_i Y_{2i} - \mu_{i+1} Y_{2i+1}, \qquad i=1,\ldots,t-1\nn\\
P_i & = & \mu_i\left(\lambda_i U_{i+1}^2 -
\lambda_{i+1} U_i^2\right), \qquad i=q+1,\ldots,q+s-t\eea
and substitute for $Y_{2i}$ in terms of $Y_{2i-1}$ and $M_i$, $i=1,\ldots,
q$;
$Y_{2i+1}$ in terms of $Y_{2i}$ and $N_i$, $i=1,\ldots,t-1$; and $U_{i+1}^2$ in
terms of $U_i^2$ and $P_i$, $i=q+1,\ldots,q+s-t$. In particular, note that
\bea\label{ron}
Y_{2t-1} = cY_1 - \sum_{i=1}^{t-1} a_i M_i + b_i N_i\eea
where $c$, $a_i$ and $b_i$ are positive constants that we do not
need to determine explicitly\footnote{These constants may be
determined by using iteratively the relations (\ref{rels}).}.
 The superpotential, in these new variables, then reads
\bea
\widetilde{W} &= & -\lambda_1 U_1^2 Y_1 + \left(\sum_{i=1}^{t-1} V_i^2 M_i
+ U_{i+1}^2 N_i\right) + V_t^2M_t + \frac{1}{\lambda_{t+1}}\mu_tU_{t+1}^2 \left(\lambda_{t}Y_{2t-1}-M_t\right)\nn\\
&+& \sum_{i=t+1}^q V_i^2 M_i - U_i^2V_i^1Y_{2i-1} + \frac{1}{\lambda_{i+1}}V_i^1U_{i+1}^2 \left(\lambda_{i}Y_{2i-1}-M_i\right) + \sum_{i=q+1}^{q+s-t}
Y_{i+q}P_i \nn\\
&+& \sum_{i=q+1+s-t}^p \lambda_i Z_i U_{i+1}^2 Y_{i+q} - \lambda_{i+1}Z_iY_{i+q}U_i^2
\eea
where one must substitute for $Y_{2t-1}$ in the first line using
(\ref{ron}). As usual, the quadratic terms lead to masses for
the corresponding fields, which must then be integrated out in the IR.
Integrating out $V_i^2$, $U_{i+1}^2$ and $Y_{i+q}$ sets
\bea
M_i &= & 0, \qquad i=1,\ldots,q\nn\\
N_i & = & 0, \qquad i=1,\ldots,t-1\nn\\
P_i & = &0, \qquad i=q+1,\ldots,q+s-t\eea
respectively.
Integrating out $M_i$, $i=1,\ldots,t-1$ sets
$V_i^2 = (a_i \lambda_t\mu_t/\lambda_{t+1}) U_{t+1}^2$. Integrating out $M_t$
sets $V_t^2 = (\mu_t/\lambda_{t+1})U_{t+1}^2$. Integrating out
$M_i$, $i=t+1,\ldots,q$ sets $\lambda_{i+1}V_i^2 = V_i^1U_{i+1}^2$.
Integrating out $N_i$, $i=1,\ldots,t-1$ sets $U_{i+1}^2 =
(b_i \lambda_t\mu_t/\lambda_{t+1}) U_{t+1}^2$. Integrating out
$P_i$, $i=q+1,\ldots,q+1+s-t$ sets $Y_{i+q}=0$.

Finally, we integrate out $U_1^2$ to obtain $\lambda_1 Y_1=\lambda_{p}
Y_{p+q}Z_p$; $Y_1$ to obtain
\bea\label{hagred}
\lambda_1U_1^2 =
(c \lambda_t\mu_t/\lambda_{t+1}) U_{t+1}^2~;\eea
and $U_{t+1}^2$ to
obtain $\mu_t\lambda_t Y_{2t-1} = \lambda_{t+1} V_{t+1}^1Y_{2t+1}$.

\begin{figure}[ht!]
\begin{center}
\epsfig{file=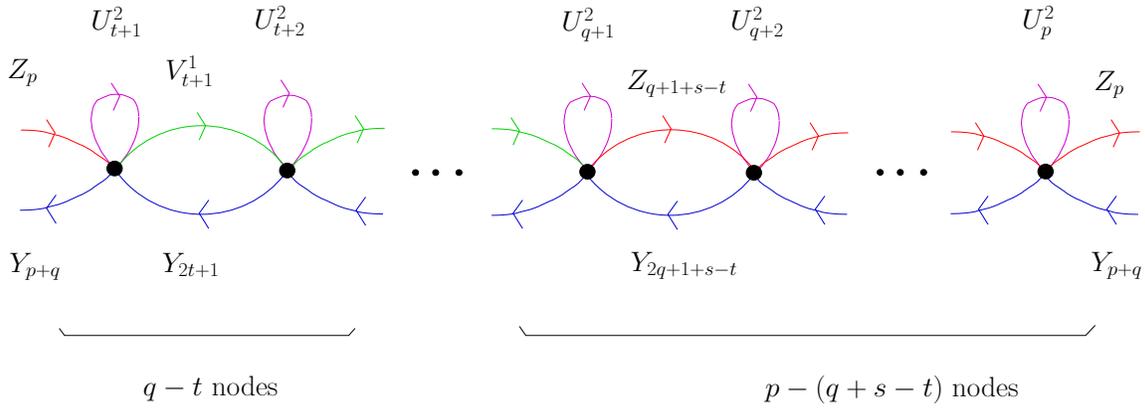,width=15.1cm,height=5.3cm}
\end{center}
\caption{Quiver for the $\mathcal{N}=2$ $A_{p-s-1}$ orbifold quiver gauge theory, obtained via
Higgsing a set of $U^1$, $V^1$ and $Z$ fields.}
\label{longquiver}
\end{figure}

All this results in the simple cubic superpotential
\bea\label{double}
\widetilde{W}_{\mathrm{eff}} &= &\sum_{i=t+1}^q \frac{\lambda_i}{\lambda_{i+1}}
V_i^1U_{i+1}^2 Y_{2i-1}
- U_i^2V_i^1 Y_{2i-1}  \nn\\
&+&\sum_{i=q+1+s-t}^p\lambda_i Z_i U_{i+1}^2 Y_{i+q} - \lambda_{i+1}Z_iY_{i+q}U_i^2~.
\eea
Here $U_{t+1}^2$ is to be identified with $U_1^2=U_{p+1}^2$  via
(\ref{hagred}), and $U_{q+1+s-t}^2$ is to be identified with
$U_{q+1}^2$  using $P_i=0$ iteratively in the relations
(\ref{rels}). As usual, the reader may check that some simple
field redefinitions effectively set all the constants in
$\widetilde{W}_{\mathrm{eff}}$ equal to one. This
is precisely the field content and superpotential of
the $\mathcal{N}=2$ $A_{p-s-1}$ orbifold theory, depicted in Figure
\ref{longquiver}.


\subsubsection{Higgsing $U^1$ and $Y$}

\begin{figure}[ht!]
\begin{center}
\epsfig{file=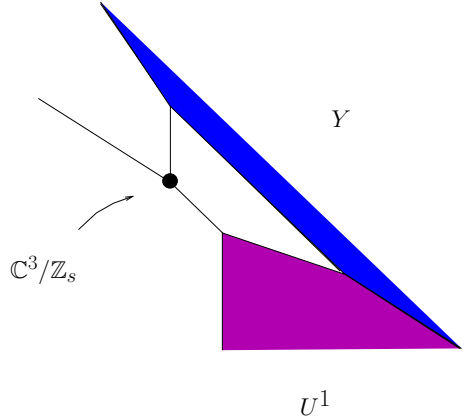,width=6cm,height=5.5cm}
\end{center}
\caption{Placing the D3-branes at the $U(1)^3$-invariant point on the exceptional
divisor, as shown, gives VEVs to a set of $U^1$ and $Y$ baryons.}
\label{UYvev}
\end{figure}

Placing the D3-branes at the corner of the exceptional divisor
$M$, as  in Figure \ref{UYvev}, implies that no $Z$ or $U^2$ fields
get non-zero VEVs. In this section we show that
a certain two-parameter
family of VEVs all flow to the same IR theory, namely
the $\mathcal{N}=2$ $A_{p-r-1}$ orbifold theory. Recall that
here $r=p-s$. This
is again precisely as expected from the gravity dual, since this
is indeed the near-horizon geometry of the stack of D3-branes.

We give the following VEVs:
\bea
U^1_i & = &\lambda_i \ \mathrm{I}_{N\times N}, \qquad i=1,\ldots,p\nn\\
Y_i & = &\mu_i\ \mathrm{I}_{N\times N}, \qquad i=1,\ldots 2t\nn\\
Y_i & = & \mu_i\ \mathrm{I}_{N\times N}, \qquad i=2q+1,\ldots,2q+r-t~.\eea
Again, this is not the most general set of VEVs we could turn on, but
rather a representative calculation.  In particular, one may also turn
on an \emph{odd} number of VEVs for the cubic $Y$ fields.
We have $0\leq t\leq r$,
with $0<r<p$, $t<q$, $r-t<p-q$.

The superpotential becomes
\bea
\widetilde{W}_{\mathrm{eff}} &= &\sum_{i=1}^t \lambda_i\mu_{2i-1}V_i^2
- \mu_{2i-1}V_i^1 U_i^2 + \mu_{2i}V_i^1 U_{i+1}^2 - \lambda_{i+1}\mu_{2i}
V_i^2\nn\\
&+& \sum_{i=t+1}^q \lambda_i V_i^2 Y_{2i-1} - U_i^2 V_i^1 Y_{2i-1}
+ V_i^1U_{i+1}^2 Y_{2i} - \lambda_{i+1}V_i^2 Y_{2i}\nn\\
&+& \sum_{i=q+1}^{q+r-t} \lambda_i\mu_{i+q}Z_i U_{i+1}^2 - \lambda_{i+1}\mu_{i+q}Z_i U_i^2
\nn\\
&+& \sum_{i=q+1+r-t}^p \lambda_i Z_i U_{i+1}^2 Y_{i+q} - \lambda_{i+1}Z_iY_{i+q}U_i^2~.
\eea
Note the linear terms in $V_i^2$ for $i=1,\ldots,2t$.
Strictly speaking we should have allowed for fluctuations of the
fields around their vacuum expectation values. These fluctuations
will give a mass to $V_i^2$, which as usual is then integrated out
in the IR. Since these fluctuation terms will turn out to be
irrelevant in the IR, we suppress them in order to keep expressions
to a manageable length. We now define
\bea\label{qels}
M_i &= & \mu_{2i}U_{i+1}^2 - \mu_{2i-1}U_i^2 \qquad i=1,\ldots,2t\nn\\
N_i & = & \lambda_i Y_{2i-1} - \lambda_{i+1} Y_{2i}, \qquad i=t+1,\ldots,q\nn\\
P_i & = & \mu_{i+q}\left(\lambda_i U_{i+1}^2 -
\lambda_{i+1} U_i^2\right), \qquad i=q+1,\ldots,q+r-t~.\eea
We then substitute for $U_{i+1}^2$ in terms of $U_i^2$ and $M_i$, $i=1,\ldots,2t$;
$Y_{2i}$ in terms of $Y_{2i-1}$ and $N_i$, $i=t+1,\ldots,q$; and
$U_{i+1}^2$ in terms of $U_i^2$ and $P_i$, $i=q+1,\ldots,q+r-t$.
Integrating out massive fields proceeds much as in the previous subsection.
In particular, however, we obtain the necessary relations
\bea
\lambda_i\mu_{2i-1} = \lambda_{i+1}\mu_{2i}, \qquad i=1,\ldots,t
\eea
on the VEVs. These effectively come
from the F-term relations. There are thus effectively only
$r$ independent VEVs for the $Y$ fields, rather than the $2t+(r-t)$
VEVs we began with. The pattern of VEVs then parallels that for the
$Z$ fields in the previous subsection.
The final effective superpotential in the IR is given by
\bea
\widetilde{W}_{\mathrm{eff}} &=& \sum_{i=t+1}^q \frac{\lambda_i}{\lambda_{i+1}} V_i^1U_{i+1}^2Y_{2i-1}
-U_i^2V_i^1Y_{2i-1}\nn\\
& +& \sum_{i=q+1+r-t}^p \lambda_i Z_i U_{i+1}^2 Y_{i+q}
- \lambda_{i+1}Z_i Y_{i+q}U_i^2~.\eea
Here $U_{t+1}^2$ is essentially identified with $U_1^2$; and $U_{q+1+r-t}^2$
is essentially identified with $U_{q+1}^2$. Note this is
precisely the same as (\ref{double}), with $r$ in place of $s$.
This is therefore the matter content and cubic superpotential
of the $A_{p-r-1}$ orbifold quiver gauge theory.


\section{Discussion}
\label{webu}

In this paper we studied deformations of SCFTs with 
Sasaki-Einstein duals, obtained by giving non-zero VEVs to baryonic operators.
We have argued that giving
expectation values to baryonic operators (and only to these) in a
superconformal quiver induces an RG flow to another IR conformal
fixed point. The supergravity backgrounds AdS/CFT dual to these
flows are \emph{warped resolved  asymptotically conical Calabi-Yau }metrics,
where the warping is induced by a stack of $N$ D3-branes placed at
some residual singularity, encoding the IR SCFT. When the geometries
and field theories are toric, one may represent the full background
in terms of pq-web-like diagrams. As explicit examples, we have
discussed the partially resolved $Y^{p,q}$ metrics presented in
\cite{np1}. The toric geometry description of the latter elucidates
the dual field theory interpretation in terms of VEVs of baryonic operators.

We have also discussed a proposal for computing the condensate
of the baryonic operators that are turned on in a given VEV-induced
RG flow. In particular, we have given further evidence for  identifying the exponentiated
on-shell Euclidean D3-brane action as the string dual to baryonic
condensates in a generic supergravity background of the above type.
This identification gives a simple sufficient condition for a
condensate to vanish, and we have checked this criterion in a number
of non-trivial examples. However, the examples studied in this paper
make clear that in a generic situation (\emph{i.e} different from
the conifold example discussed in \cite{KM}) the calculation of the
condensate that we have outlined is necessarily rather more
complicated. Indeed, recall that the AdS/CFT definition of a
baryonic particle involves specifying a supersymmetric 3-submanifold
\emph{and} a flat (hence torsion) line bundle.
Incorporating this into the instantonic D3-brane calculation
requires studying the extension of this pair of data from the
boundary to the interior. In turn, this requires a careful analysis
of the flat background fields in a given geometry. 
These issues will be addressed in future work \cite{np3}.


\subsection*{Acknowledgments}
\noindent We would like to thank I. Klebanov, J. Maldacena,
A. Murugan, Y. Tachikawa  and S.-T. Yau for useful
discussions. The research of J. F. S. at the University of Oxford is
supported by a Royal Society University Research Fellowship, although the majority
of this work was carried out at Harvard University, supported by NSF grants DMS-0244464,
DMS-0074329 and DMS-9803347. D. M. would like to thank the physics
and mathematics departments of Harvard University for hospitality
during the early stages of this work. He acknowledges support from
NSF grant PHY-0503584.


\end{document}